\documentclass[showpacs,showkeys,preprintnumbers,nofootinbib]{revtex4}
\usepackage{graphicx}
\usepackage{amsmath}
\usepackage{enumerate}

\DeclareMathOperator{\Tr}{Tr}

\DeclareMathOperator{\su}{su}
\DeclareMathOperator{\U}{U}
\DeclareMathOperator{\SU}{SU}

\newcommand{\be}{\begin{equation}}
\newcommand{\ee}{\end{equation}}
\newcommand{\bea}{\begin{eqnarray}}
\newcommand{\eea}{\end{eqnarray}}

\def\Xint#1{\mathchoice
   {\XXint\displaystyle\textstyle{#1}}%
   {\XXint\textstyle\scriptstyle{#1}}%
   {\XXint\scriptstyle\scriptscriptstyle{#1}}%
   {\XXint\scriptscriptstyle\scriptscriptstyle{#1}}%
   \!\int}
\def\XXint#1#2#3{{\setbox0=\hbox{$#1{#2#3}{\int}$}
     \vcenter{\hbox{$#2#3$}}\kern-.5\wd0}}

\def\dashint{\Xint-}

\begin{document}
\title{Weak-coupling analysis of the single-site large-N gauge theory
  coupled to adjoint fermions}
\date{\today}
\author{Robert Lohmayer}
\email{robert.lohmayer@fiu.edu}
\author{Rajamani Narayanan}
\email{rajamani.narayanan@fiu.edu}
\affiliation{ 
Department of Physics, Florida International University,
Miami, FL 33199, USA.}

\begin{abstract}
We consider the leading-order expression at weak coupling for a single-site
 large-$N$ gauge theory coupled to adjoint fermions. We study the
case
of overlap and Wilson fermions. We extend the theory to real values of
the number of fermion flavors and restrict ourselves to asymptotically
free
theories.
Using a four-dimensional density
function
for the distribution of the eigenvalues of the link variables, we show
that
it is possible to recover the infinite-volume continuum limit for a
certain range of fermion flavors if we use fermions with a bare mass
of zero.
Our use of the four-dimensional density function is supported by a
direct
analysis of the lattice action.
\end{abstract}


\preprint{\today}

\maketitle

\section{Introduction}
Nonabelian gauge field theories coupled to fermions in some
representation of the gauge group are asymptotically free as long as
the number of fermion flavors is less than a certain number. Within
this allowed range of fermion flavors, the theory is expected to be
confining in some range at the lower end and it is expected to be
conformal
at the higher end. Identification of the critical number of fermion
flavors
that separate the confining region from the conformal region is a
non-perturbative task that has recently received considerable attention within
the
lattice field theory
community \cite{Hasenfratz:2013uha}--\cite{Appelquist:2007hu}.
Several issues need to be resolved before such an endeavor can make
some
physics conclusions. These include (a) How does one deal with a
conformal theory on a lattice; (b) How does one conclude that a
certain model exhibits features of near conformal features; (c) How
does
one compute the location of the infra-red fixed point in a lattice
model.
Since finite-volume effects need to be understood carefully and since
one has to be close to the chiral limit to understand the above
issues, the numerical simulations are inherently large scale in
nature.

An attractive alternative has been proposed to study large-$N$ gauge
field theories coupled to adjoint fermions on a single-site lattice.
It has been argued that reduction to a single-site lattice should hold
in such theories~\cite{Kovtun:2007py} and tested numerically using
a variety of methods to see if one can reproduce the continuum
infinite-volume theory by working on a single-site lattice
\cite{Bedaque:2009md}--\cite{Gonzalez-Arroyo:2013gpa}.
With the exception of
\cite{Gonzalez-Arroyo:2013gpa,GonzalezArroyo:2012st},
all attempts have considered the Eguchi-Kawai reduction and 
numerically argued that the single-site theory is in the correct
continuum phase. Asymptotic freedom is maintained in these theories
if the number of Dirac flavors is less than $\frac{11}{4}$. 
Two-loop perturbative beta function would suggest the existence of an
infra-red fixed point if the the number of fermion flavors is greater
than $\frac{17}{16}$. With this in perspective, the single-site model
with one massless adjoint overlap-Dirac fermion was extensively
studied
in~\cite{Hietanen:2012ma}. Numerical results suggest that the coupling
runs much faster than what is predicted by continuum two-loop
perturbation theory at the lattice couplings that were considered. 
In order to better understand the connection between single-site
lattice models and infinite-volume continuum theories, we decided to
revisit the problem of perturbation theory on the single-site lattice
in this paper.

We will consider the weak-coupling limit and the only parameters we
will
consider are the number of fermion flavors which we will extend to
take
on all real values in the range $\left[0,\frac{11}{4}\right]$
and the fermion mass.
The main aim of this paper is to use a four-dimensional density
function to answer two questions:
\begin{enumerate}
\item What is the range of fermion flavors for which the single-site
massless theory can be expected to reproduce the infinite-volume
continuum theory?
\item Can we reproduce the infinite-volume continuum theory with
  massive fermions?
\end{enumerate}
We will provide an answer to both these questions using Wilson
fermions
and overlap fermions. We will not consider the case of twisted
reduction in this paper.

\section{The single-site model}

The single-site partition function for a $\SU(N)$ gauge theory coupled to
$f$ flavors of fermions in the adjoint representation is  
\begin{align}
Z = \int \prod_\mu dU_\mu e^{S_g + f S_f}
\end{align}
with Haar measure $dU$. The Wilson gauge action
is
\begin{align}
S_g = bN \sum_{\mu , \nu=1}^d \Tr \left[ U_\mu U_\nu U_\mu^\dagger
U_\nu^\dagger -1\right].
\end{align}
The fermion action is
\begin{align}
S_f = \ln\det H_{w,o}
\end{align}
with the subscript $w$ for Wilson fermions and $o$ for overlap
fermions.
The Hermitian Wilson Dirac operator 
for massive adjoint fermions is
given by
\begin{align}
H_w(m_w) = \begin{pmatrix} 4 + m_w -\frac{1}{2}\sum_\mu \left( A_\mu + A_\mu^t\right)
& \frac{1}{2}\sum_\mu \sigma_\mu \left(A_\mu - A_\mu^t\right) \cr
-\frac{1}{2}\sum_\mu \sigma^\dagger_\mu \left(A_\mu - A_\mu^t\right) &
-4 - m_w+\frac{1}{2}\sum_\mu \left( A_\mu + A_\mu^t\right)\cr
\end{pmatrix}\,,
\label{wilson}
\end{align}
where $m_w$ is the bare Wilson fermion mass.
The adjoint gauge fields are given by
\begin{align}
A_\mu^{ab} = \frac{1}{2}\Tr \left [ T^a U_\mu T^b U^\dagger_\mu\right],\label{adjoint}
\end{align}
where
$T^a$, $a=1,\ldots, (N^2-1)$ are traceless Hermitian matrices
that generate the $\su(N)$ Lie algebra and satisfy
\begin{align}
\Tr T^a T^b = 2\delta^{ab}\,;\ \ \ \ 
[T^a, T^b] = \sum_c i f^{ab}_c T^c.\label{structure}
\end{align}
The Hermitian  
massive overlap Dirac operator is defined by
\begin{align} 
H_o(m_o) = \frac{1}{2}\left [ \left( 1 + m_o \right)\gamma_5 +
\left(1-m_o\right)\epsilon\left[H_w(m_w)\right]\right]\,,\label{hover}
\end{align} 
where $m_o\in[0,1]$ is the bare overlap fermion mass and
$m_w < 0$ is the irrelevant Wilson mass parameter.

The total action depends on $d$ $\SU(N)$ matrices and the gauge transformation is
\begin{align}
U_\mu \to g U_\mu g^\dagger.
\end{align}
Note that the eigenvalues of $U_\mu$ are gauge invariant.
We cannot fix a gauge such that one of the $U_\mu=1$ since we are on a
single-site
lattice. The action has an additional $\U^d(1)$ symmetry given by
\begin{align}
U_\mu \to e^{i\alpha_\mu} U_\mu\label{znsymm}
\end{align}
with $ 0 \leq \alpha_\mu < 2\pi$. Restricting $\alpha_\mu$ to
$\frac{2\pi k_\mu}{N}$ with integers $0 \le k_\mu < N$ keeps it in
$\SU(N)$; otherwise we have trivially extended the $\SU(N)$ theory to a
$\U(N)$ theory.
The four Polyakov loop operators, given by
\begin{align}
P_\mu = \Tr U_\mu \,,
\end{align}
are gauge invariant but not invariant under (\ref{znsymm}). 
If the $\U^d(1)$ symmetry is not broken, then the eigenvalues of
all $U_\mu$ are uniformly distributed on the unit circle and $P_\mu=0$
(the reverse statement is not necessarily true because the eigenvalues
in different directions might be correlated). In the following, we set the number of Euclidean space-time dimensions $d$ to $4$.

\section{Leading-order perturbation theory and the density function}

The symmetry given by (\ref{znsymm}) is spontaneously broken in the
weak-coupling limit if we do not have adjoint fermions even when there
are a finite number of flavors of fundamental fermions~\cite{Bhanot:1982sh}.
We want to study if this symmetry is spontaneously broken
in the weak-coupling limit in the presence of adjoint fermions.
 We set
\begin{align}
U_\mu = V_\mu D_\mu V^\dagger_\mu\,;\ \ \ \  
D_\mu^{jk} = e^{i\theta^j_\mu}\delta^{jk},\label{upert}
\end{align}
and expand around $V_\mu=1$
to compute observables in perturbation
theory.

The expression for the partition function at leading order at weak coupling is known~\cite{Hietanen:2009ex} and is given by
\begin{align}
Z^0 &= \int  \left[ \prod_\mu \prod_i d\theta^i_\mu \right] e^{ S^0}\,;
&S^0&= S_g^0 + f S_f^0\,;   \cr    
S_g^0&=-\sum_{i\ne j} \ln \hat p^{ij}\,;
&\hat p^{ij}& = \sum_\mu 4 \sin^2 \frac{\theta_\mu^i-\theta_\mu^j}{2}\,.\label{zquad}
\end{align}
The fermionic contribution is
\begin{align}
S_{w,o} ^0 =  2 \sum_{i \ne j} \ln
\gamma_{w,o}^{ij}(m_{w,o})\,,\label{ftree}
\end{align}
where we have removed a $\theta$-independent term that arises from the
zero modes for massless fermions and assumed that 
$\hat p^{ij} \ne 0$ if $i\ne j$.
The non-zero modes are given by 
\begin{align}
\gamma_w^{ij}(m_w)&=\left(m_w+\frac{\hat p^{ij}}{2} \right)^2 + \bar p^{ij} \,; \qquad\qquad
\bar p^{ij} =  \sum_\mu \sin^2 \left(\theta_\mu^i-\theta_\mu^j\right)\,; \cr
\gamma_o^{ij}(m_o,m_w) & = \frac{1+m_o^2}{2}+\frac{1-m_o^2}{2} \frac{m_w+\frac{\hat
    p^{ij}}{2} }{\sqrt{\gamma_w^{ij}(m_w)}}\, .
\end{align}
Owing to the  symmetry given by  (\ref{znsymm})
$S^0$  is invariant under $\theta_\mu^i\to\theta_\mu^i+\alpha_\mu$ for any choice of $\alpha_\mu$. 

As $N\to\infty$, we assume that we can define a joint distribution,
$\rho(\theta)$, in the following sense:
At any finite $N$, for a fixed choice of $\theta_\mu^i$, $i=1,\ldots,N$ and
$\mu=1,\ldots,4$, let
\begin{align}
\rho(\theta) = \frac{1}{N} \sum_i \prod_\mu
\delta(\theta_\mu-\theta_\mu^i)\,;\qquad\qquad \int \prod_\mu d\theta_\mu \rho(\theta) =1\,,\label{distdef}
\end{align}
where $\delta$ denotes the $2\pi$-periodized delta function normalized to $\int_{-\pi}^\pi d\theta \delta(\theta)=1$.
We can then rewrite $S^0$ in (\ref{zquad}) as
\begin{align}
S^0_{g,f} &= N^2 
\dashint d^4\theta d^4\phi\,
\rho(\theta) S_{g,f}(\theta-\phi) \rho(\phi)\,;
\cr
S_g(\theta) &= -\ln \hat p \,; \qquad\hat p =
\sum_\mu 4 \sin^2 \frac{\theta_\mu}{2}\,;\cr
S_f(\theta) &= 2 \ln \gamma_{w,o}(m_{w,o})\,; \cr
\gamma_w (m_w)&=\left(m_w+\frac{\hat p}{2} \right)^2 + \bar p\,  ;  \qquad
\bar p  =  \sum_\mu \sin^2 \theta_\mu\,;
\cr
\gamma_o(m_o,m_w) & = \frac{1+m_o^2}{2}+\frac{1-m_o^2}{2} \frac{m_w+\frac{\hat
    p}{2} }{\sqrt{\gamma_w(m_w)}}\, . \label{dentree}
\end{align}
Since there is a restriction  in the sum that appears in (\ref{zquad})
and (\ref{ftree}), we have to evaluate the principal value of the
integral appearing in (\ref{dentree}) by excluding a small region
around $\theta=\phi$. 
The integral, $\dashint$, indicates the Cauchy Principal Value.
Finally, we can write
\begin{align}\label{eq:Z[rho]}
Z^0 = \int [d\rho] e^{S^0}\,; \qquad\qquad S^0=S_g^0 + f S_f^0\,,
\end{align}
where by $\int [d\rho]$ we mean the integral over all possible
choices for $\theta_\mu^j$, $j=1,\ldots,N$ and $\mu=1,\ldots,4$.

We now assume that, as $N\to\infty$, the integral in \eqref{eq:Z[rho]} will be dominated by a single distribution $\rho(\theta)$, maximizing $S^0[\rho]$.
We will only allow distributions that are non-negative everywhere with
the normalization condition in (\ref{distdef}). Furthermore, we assume
that the dominating distribution $\rho(\theta)$ is smooth and finite for all $\theta$ (in contrast to $\rho$ defined in \eqref{distdef} for angle configurations at finite $N$).
Since the singular nature of $S(\theta)$ in (\ref{dentree}) is only logarithmic\footnote{A special case are massless fermions at $f=1/2$, for which $S_g(\theta)+f S_f(\theta)$ is finite at $\theta=0$.}, the integrals are then finite even if we drop the principal-value restriction $\theta\neq\phi$.
Clearly, $S^0$ in (\ref{dentree}) is invariant under $\rho(\theta) \to
\rho(\theta+\alpha)$ for any choice of $\alpha$, corresponding
to the invariance under  (\ref{znsymm}).
  
Owing to the periodic and symmetric nature of $S_{g,f}(\theta)$, it follows that
\begin{align}
&\int_{-\pi}^\pi \prod_\nu \frac{d\phi_\nu}{2\pi}\, S_{g,f}(\theta-\phi)\, e^{i \sum_\mu k_\mu \phi_\mu}
= \lambda^{(g,f)}_{k}\, e^{i \sum_\mu k_\mu \theta_\mu}\,;\label{eigen}\\
\lambda^{(g,f)}_k = &\int_0^\pi \prod_\nu \frac{d\phi_\nu}{\pi}\,
S_{g,f}(\phi) \prod_\mu \cos(k_\mu\phi_\mu)\,. 
\label{eigen2}
\end{align}
Therefore, Fourier expanding
\begin{align}\label{eq:FourierExpansion}
\rho(\theta)= \frac1{(2\pi)^4} \sum_k c_k e^{i \sum_\mu k_\mu \theta_\mu} \qquad\textnormal{with}\qquad c_{-k}=c_k^\ast\,,\quad c_0=1
\end{align}
results in 
\begin{align}\label{eq:S-ck}
S^0_{g,f}=N^2 \sum_k c_kc_k^\ast \lambda_k^{(g,f)}\,,
\end{align}
provided $\rho(\theta)$ is such that we can interchange the order of principal-value integration and sums over Fourier modes when we insert \eqref{eq:FourierExpansion} in \eqref{dentree}. (If this is not the case, e.g. if $\rho(\theta)$ is of the form \eqref{distdef}, we expect the infinite sum in \eqref{eq:S-ck} to be diverging.)

If all the eigenvalues, 
\begin{align}
\lambda_k=\lambda^{(g)}_k + f \lambda^{(f)}_k
\end{align} 
for $k\ne 0$ are smaller than zero,
the constant mode, $\rho(\theta) = \frac{1}{(2\pi)^4}$, will dominate
in the large-$N$ limit (i.e., $c_k\to 0$ for $k\neq0$) and the single-site model will be in the
correct continuum phase and possibly reproduce the infinite-volume
continuum theory. 
In the next section, we will obtain the region in the $(f,m_w)$ plane for Wilson fermions and
in the $(f,m_o,m_w)$ space for overlap fermions where this is the case.
Focusing on certain points in the allowed space we will compare the
infinite-$N$ action from (\ref{dentree}) with a numerically obtained
maximum of the finite-$N$ action in (\ref{zquad}) to get a feel for
the size of the finite-$N$ effects.

If some of the eigenvalues are larger than zero, then the action $S^0$
in (\ref{dentree}) will not be maximized by $\rho(\theta) =
\frac{1}{(2\pi)^4}$ and some $c_k$ ($k\neq0$) will be non-zero. Since
the action in \eqref{eq:S-ck} is quadratic, the maximum will be
obtained at the boundary of the domain of allowed values for the
$c_k$'s, which is determined by the condition $\rho(\theta)\geq0$ for
all $\theta$. Therefore, $S[\rho]$ will be maximized by a
$\rho(\theta)$ which is zero at least at one point in
the four-dimensional Brillouin zone. Due to the shift-invariance, there will then be a
class of densities, related by $\rho(\theta)\to\rho(\theta+\alpha)$
with arbitrary $\alpha$, having identical maximum action resulting in
a spontaneous breaking of the $\U^d(1)$ symmetry in (\ref{znsymm}).

\section{Investigation of the allowed regions}

\subsection{Overlap fermions}

We will start with the action for $S^0$ as given in (\ref{dentree})
and find the eigenvalues $\lambda_{k}$ defined in (\ref{eigen}) for
all $k_\mu \leq 7$ ($\lambda_k$ is invariant under sign changes and permutations of the $k_\mu$). In the following, we consider only $k\neq0$.
In order to compute
the eigenvalues, we need to perform the integral in (\ref{eigen2})
numerically
and we will do this using a four-dimensional uniform Riemann sum.

\begin{figure}[ht]
\centerline{
\includegraphics[width=140mm]{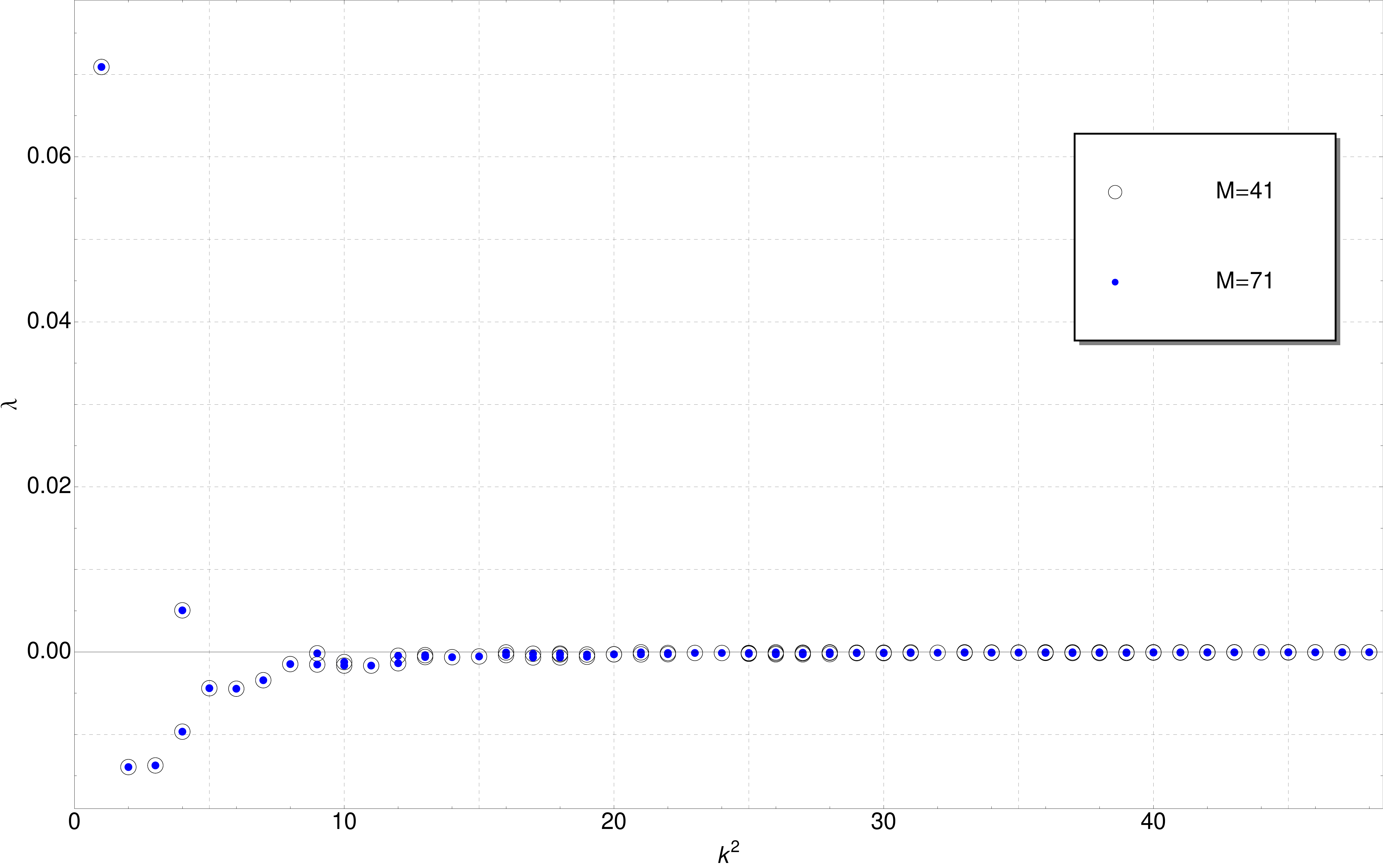}
}\caption{
Eigenvalues $\lambda_k = \lambda_k^{(g)} + f \lambda_k^{(o)}$ as a function of $k^2$ for the massless overlap Dirac operator with $f=1$ and
$m_w=-1$ obtained using numerical integration with $M^4$ equally
spaced points in the four-dimensional integration space.
} \label{fig1}
\end{figure}

A sample plot is shown in Fig.~\ref{fig1} where we have computed the
eigenvalues $\lambda_k=\lambda_k^{(g)}+f\lambda_k^{(f)}$ for massless overlap fermions with $f=1$ and $m_w=-1$.
The results are obtained with $M^4$ equally spaced points in the 
four-dimensional integration space and we used $M=41$ and $M=71$ to
show that we have reached the limit of the continuum integral.
Since two eigenvalues are positive, $(f=1,m_o=0,m_w=-1)$ is not a point in the allowed region for overlap fermions.

\begin{figure}[ht]
\centerline{
\includegraphics[width=140mm]{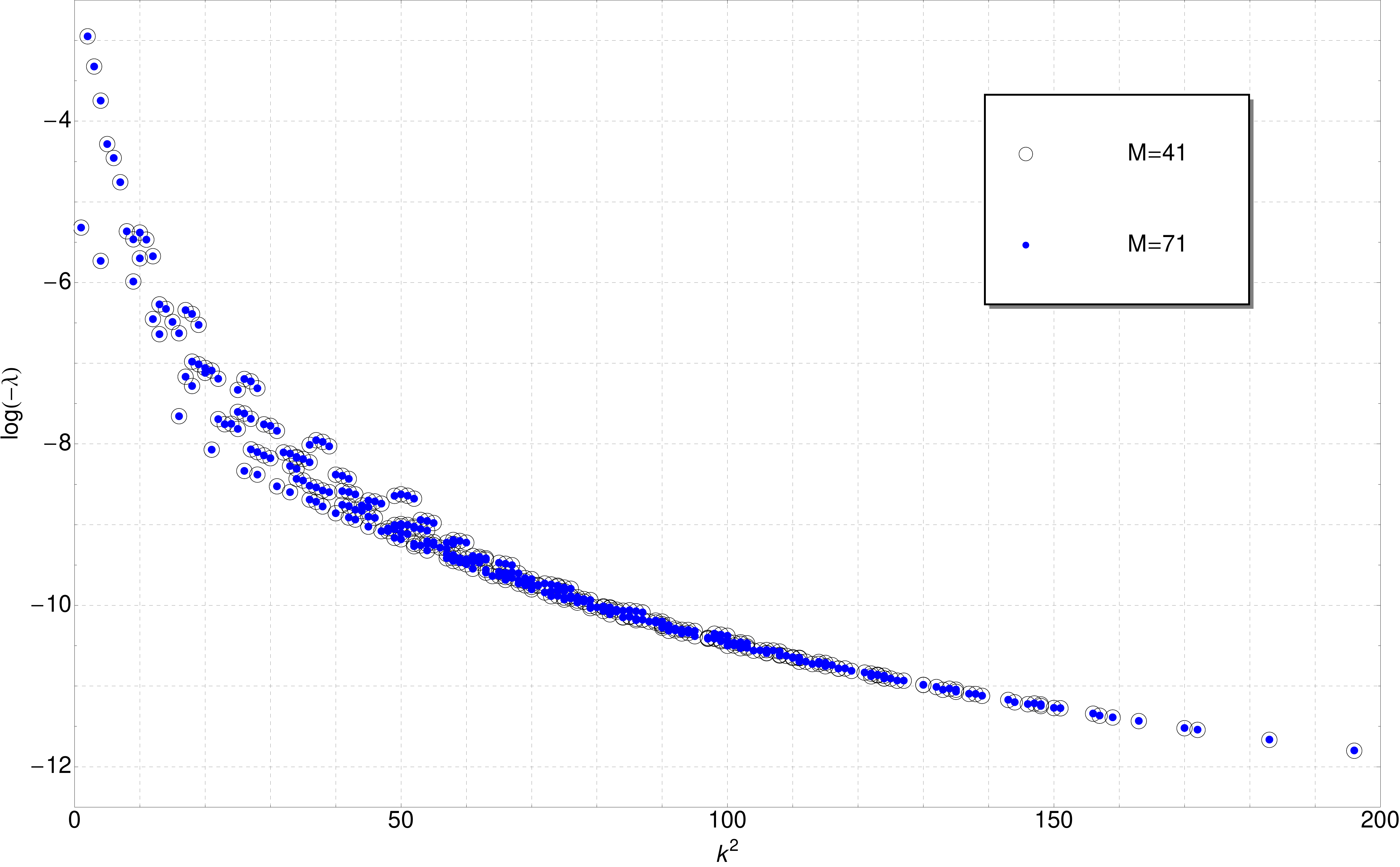}
}\caption{
Logarithm of the eigenvalues for the massless overlap Dirac operator with $f=2$ and
$m_w=-1$ obtained using numerical integration with $M^4$ equally
spaced points in the four-dimensional integration space.
} \label{fig2}
\end{figure}

As a second example, we set $f=2$, keeping $m_o=0$ and $m_w=-1$. In this case, we find all eigenvalues $\lambda_k$ to be negative, making this a point inside the allowed region. In Fig.~\ref{fig2}, we have plotted $\ln(-\lambda_k)$ as a function of $k^2$ to show that even in the log-scale we have a good estimate for the continuum integral.

Numerically, we find that $\lambda_k^{(g)}>0$ for all $k$, which means that a point $(f,m_o,m_w)$ will be inside the allowed region (defined by $\lambda_k=\lambda_k^{(g)}+f\lambda_k^{(f)}<0$ for all $k$) iff
\begin{enumerate}[(i)]
\item  $\lambda_k^{(o)}(m_o,m_w)<0$ for all $k$,
\item $f>\max_k \left\{ - \lambda^{(g)}_k/\lambda^{(o)}_k \right\}$.
\end{enumerate}      
We observe that the eigenvalues $\lambda_k^{(g)}$ and $\lambda_k^{(o)}$ go to zero as $k\to\infty$. As they approach zero from opposite sides ($\lambda^{(g)}>0$, $\lambda^{(o)}<0$), we potentially have to consider all $k$ in order to be able to determine the boundary of the allowed region in the $(f,m_o,m_w)$ space.

\begin{figure}[ht]
\centerline{
\includegraphics[width=140mm]{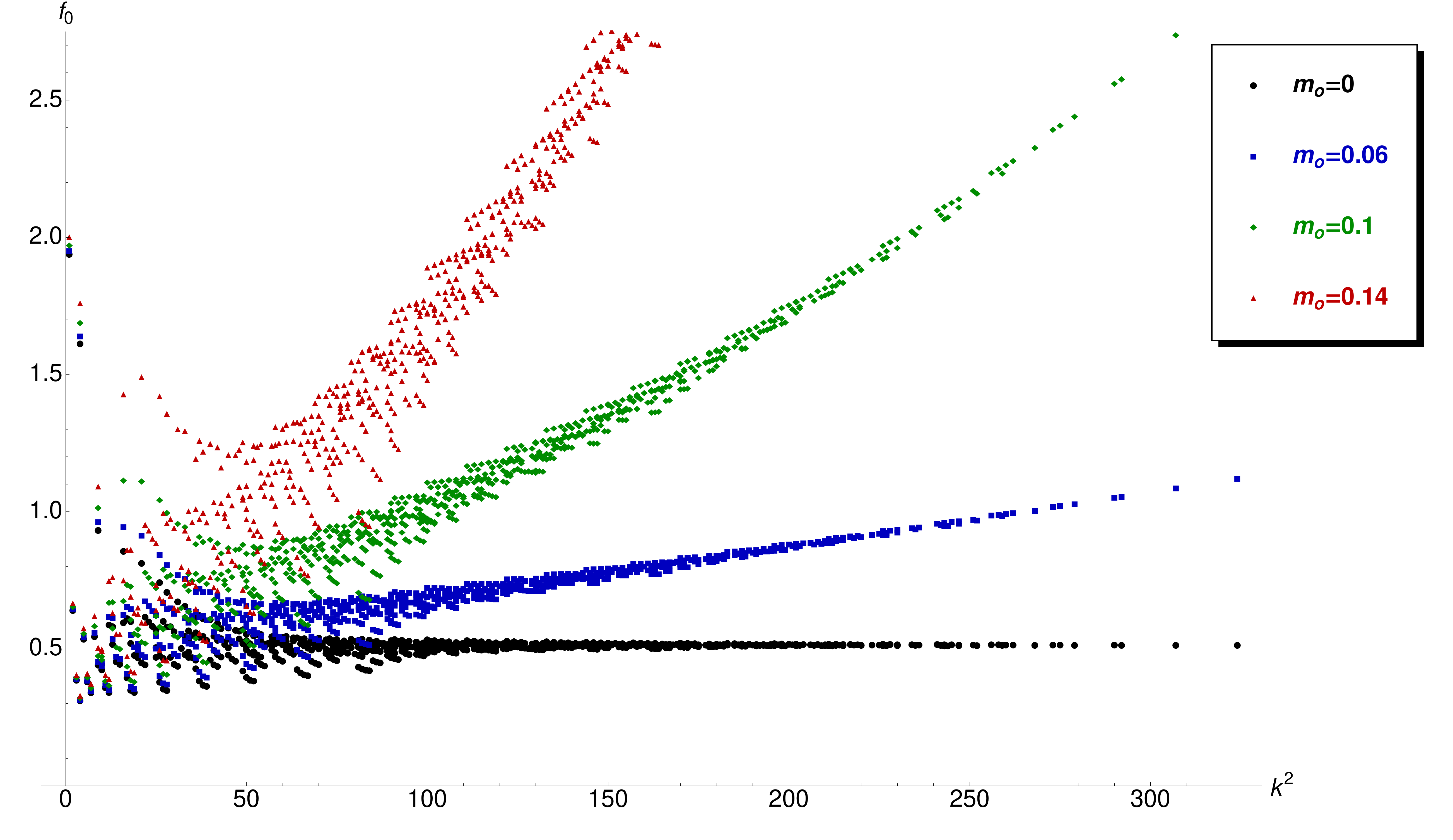}
}\caption{
Plots of $f_0\equiv - \lambda^{(g)}_k/\lambda^{(o)}_k$ (for $k_\mu\leq 9$) at $m_w=-1$ and different choices for $m_o$. For $m_o=0$, $f_0\to0.5$ as $k^2\to\infty$; for all $m_o>0$, $f_0\to\infty$ as $k^2\to\infty$. The boundary of the allowed region is determined by $\max_k f_0(k)$.
} \label{fig:overlap-f-vs-mo}
\end{figure}

\begin{figure}[ht]
\centerline{
\includegraphics[width=140mm]{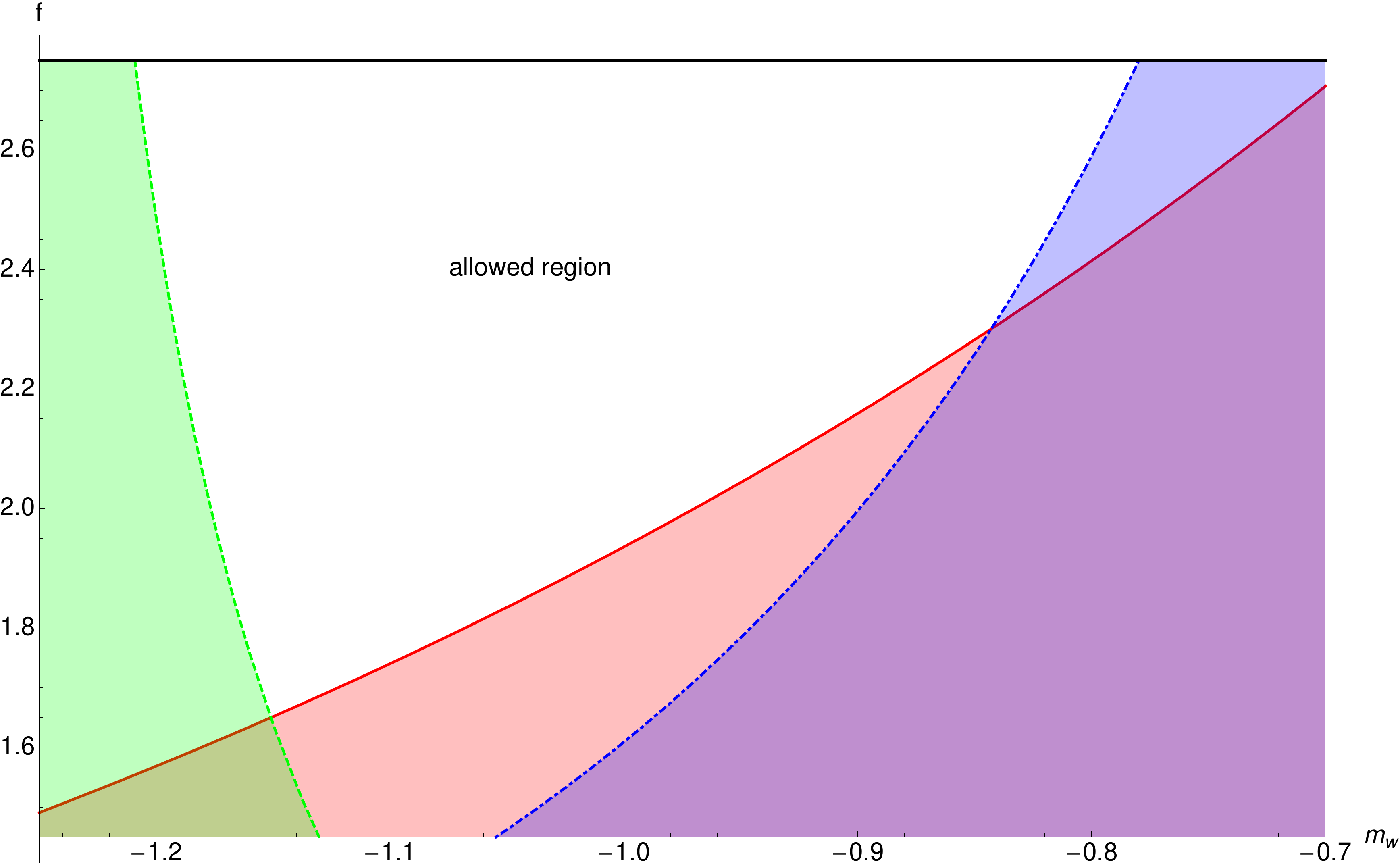}
}\caption{
Boundary of the allowed region in the $(m_w,f)$-plane for massless overlap fermions ($m_o=0$). The different lines show $- \lambda^{(g)}_k/\lambda^{(o)}_k$ for $k=(2,2,2,2)$ (green, dashed), $k=(1,0,0,0)$ (red, solid), and $k=(2,0,0,0)$ (blue, dot-dashed). The intersection points are at $(m_w,f)\approx(-0.843,2.30)$ and $(m_w,f)\approx(-1.15,1.65)$.  
} \label{fig:allowed}
\end{figure}

Let us first consider the case $m_o=0$. As $k\to\infty$, the integrals determining the eigenvalues in \eqref{eigen2} are dominated by $\phi_\nu\in(0,2\pi/k_\nu]$ since both $S_g$ and $S_o$ diverge as $\phi\to0$. Furthermore, expanding $S(\phi)$ around $\phi=0$, we obtain
\begin{align}
S_g(\phi)=-\ln(\phi^2)+\ldots\,; \qquad\qquad S_o(\phi)=2 \ln(\phi^2)+\ldots\,,
\end{align}  
indicating that $- \lambda^{(g)}_k/\lambda^{(o)}_k\to \frac 12$ as $k\to\infty$ for all $m_w<0$. Computing the eigenvalues numerically, we indeed find that $- \lambda^{(g)}_k/\lambda^{(o)}_k$ rapidly converges to $\frac 12$ for large $k$ (cf.~Fig.~\ref{fig:overlap-f-vs-mo} for an example). Therefore, for $m_o=0$, the allowed region in the $(m_w,f)$-plane is determined by eigenvalues $\lambda_k$ with $k$ being small (cf.~Fig.~\ref{fig:overlap-f-vs-mo}). Considering only $f\leq \frac{11}4$, we find numerically that the maximum $- \lambda^{(g)}_k/\lambda^{(o)}_k$, which leads to the boundary of the allowed region, is obtained at $k=(2,2,2,2)$ for $m_w\in[-1.21,-1.15]$, at $k=(1,0,0,0)$ for $m_w\in[-1.15,-0.843]$, and at $k=(2,0,0,0)$ for $m_w\in[-0.843,-0.780]$. $m_w\notin [-1.21,-0.780]$ is not allowed. For a plot of the boundary of the allowed region in the $(m_w,f)$-plane for $m_o=0$ see Fig.~\ref{fig:allowed}.

\begin{figure}[ht]
\centerline{
\includegraphics[width=130mm]{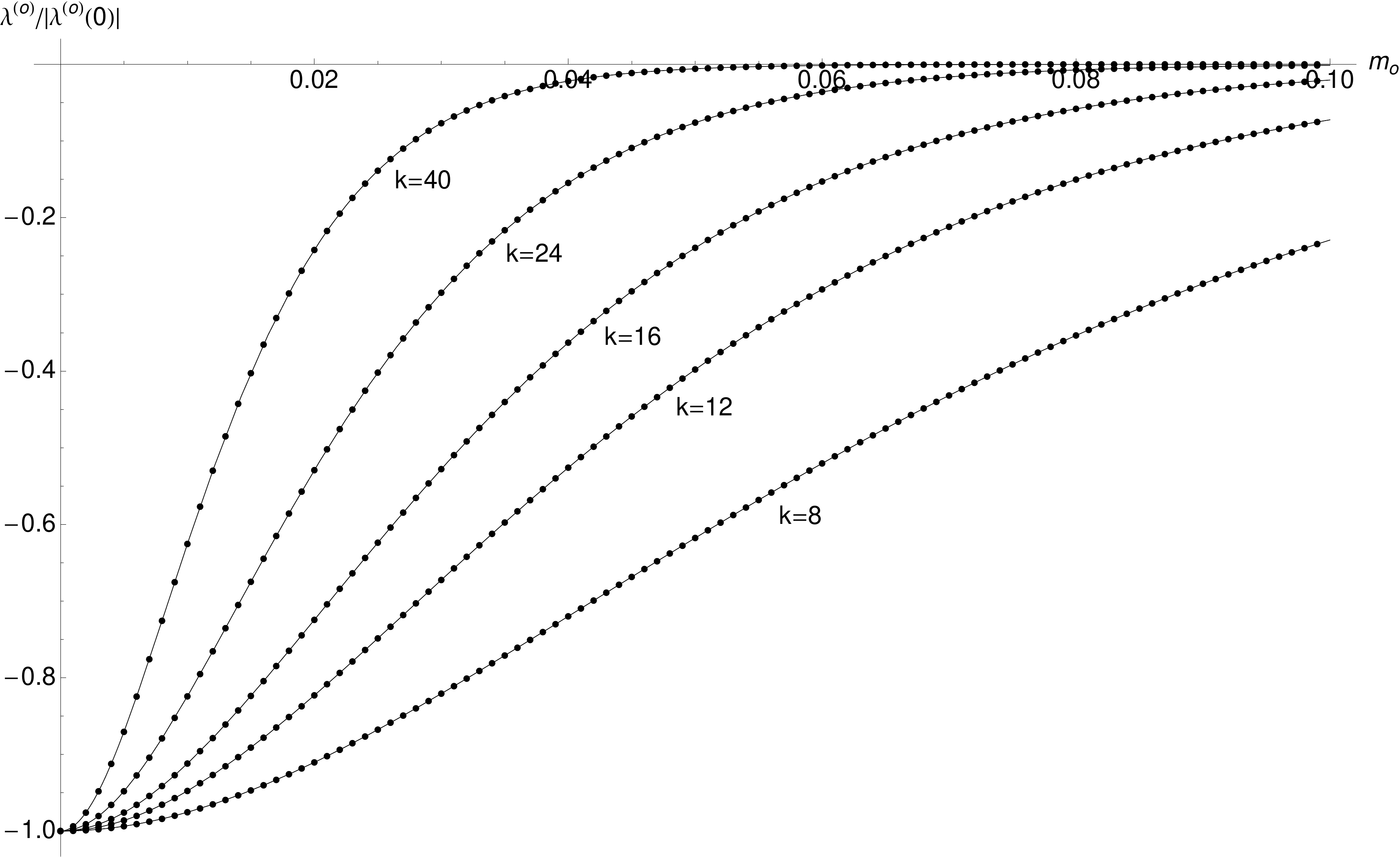}
}\caption{
Plots of $\lambda_k^{(o)}(m_o)/|\lambda_k^{(o)}(m_o=0)|$ at $m_w=-1$
for modes with $k_\mu=k\ \forall \mu$ and different choices of
$k$. The ratio goes to zero as $k\to\infty$ for every $m_o>0$, with
the rate of convergence increasing with $m_o$.
} \label{fig:ov-highk}
\end{figure}

For $m_o>0$, the divergence of $S_o(\phi)$ at $\phi=0$ is regulated and therefore $\lambda_k^{(o)}(m_o)/\lambda_k^{(o)}(m_o=0)\to 0$ as $k\to\infty$. Fig.~\ref{fig:ov-highk} shows some examples for the dependence of $\lambda_k^{(o)}(m_o)$ on $m_o$ and $k$ at $m_w=-1$. 
Together with our results for the massless case, this immediately
implies that $- \lambda^{(g)}_k/\lambda^{(o)}_k\to\infty$ as
$k\to\infty$ (see Fig.~\ref{fig:overlap-f-vs-mo} for numerical
results). 
Therefore, it is necessary to keep $m_0=0$ in the weak-coupling limit.
 
\subsection{Wilson fermions}
\begin{figure}[ht]
\centerline{
\includegraphics[width=140mm]{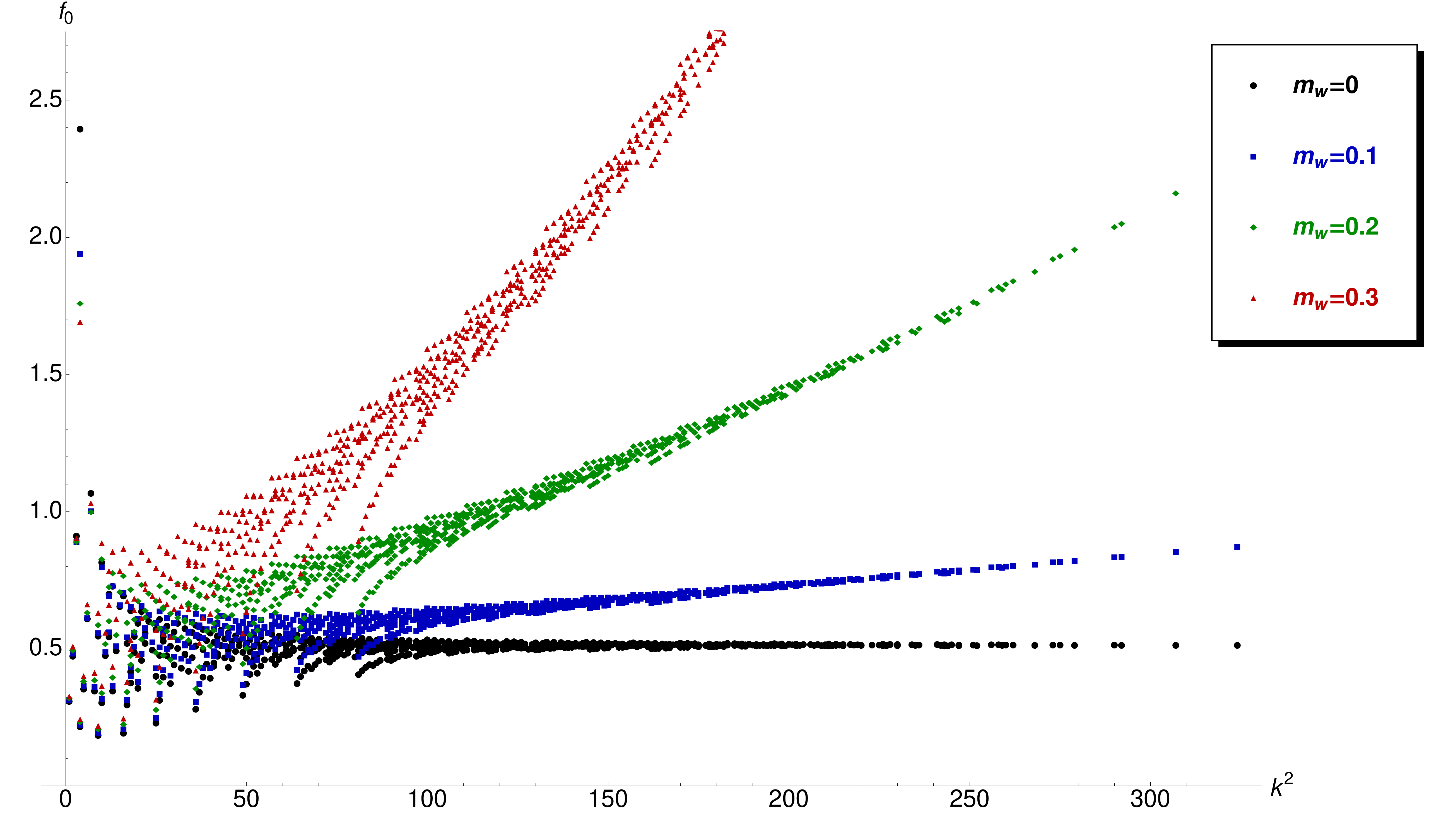}
}\caption{
Plots of $f_0\equiv - \lambda^{(g)}_k/\lambda^{(w)}_k$ (for $k_\mu\leq 9$) for different choices of $m_w$. For $m_w=0$, $f_0\to0.5$ as $k^2\to\infty$; for $m_w>0$, $f_0\to\infty$ as $k^2\to\infty$.
} \label{fig:wilson-f-vs-mw}
\end{figure}

\begin{figure}[ht]
\centerline{
\includegraphics[width=140mm]{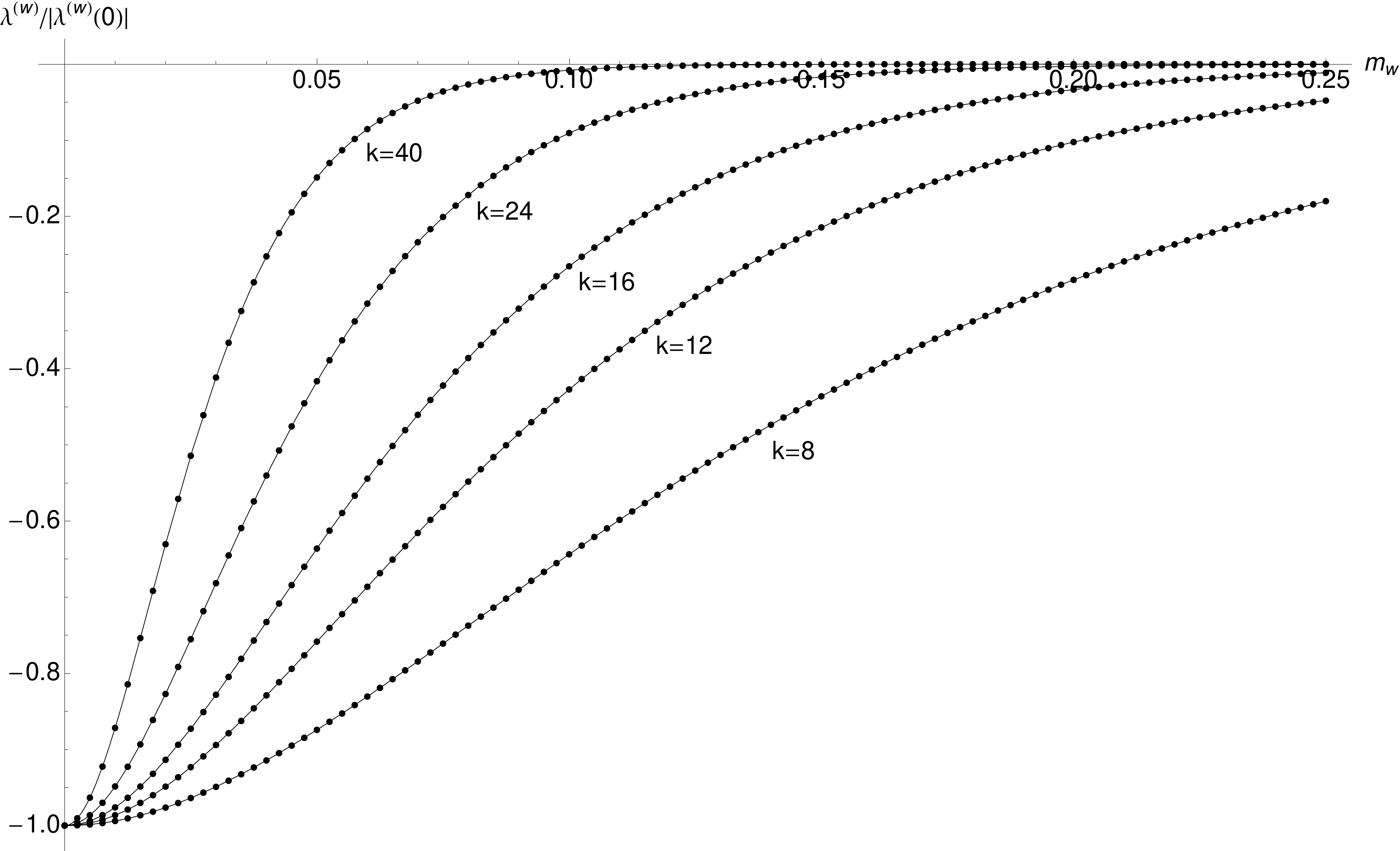}
}\caption{
Plots of $\lambda_k^{(w)}(m_w)/|\lambda_k^{(w)}(m_w=0)|$ for modes with $k_\mu=k\ \forall \mu$ and different choices of $k$.
} \label{fig:wilson-highk}
\end{figure}

The scenario for Wilson fermions is very similar to the overlap case described above, with $m_w$ now playing the role of $m_o$ and no additional irrelevant parameter. 
For $m_w=0$, we have $S_w(\phi)=2 \ln(\phi^2)+\ldots$ for small $\phi$, and therefore $- \lambda^{(g)}_k/\lambda^{(w)}_k\to \frac 12$ as $k\to\infty$, while for $m_w>0$, we find $- \lambda^{(g)}_k/\lambda^{(w)}_k\to \infty$ as $k\to\infty$, which means that $m_w>0$ is not allowed. Some numerical results are shown in Figs.~\ref{fig:wilson-f-vs-mw} and \ref{fig:wilson-highk}, which directly correspond to Figs.~\ref{fig:overlap-f-vs-mo} and \ref{fig:ov-highk} for the overlap case.
For $m_w=0$, the maximum $- \lambda^{(g)}_k/\lambda^{(w)}_k$ is
obtained at $k=(1,1,1,1)$, for which $-
\lambda^{(g)}_k/\lambda^{(w)}_k=2.39$
(cf.~Fig.~\ref{fig:wilson-f-vs-mw}). Therefore, only $m_w=0$ and
$f>2.39$ is allowed for Wilson fermions in the weak-coupling limit.

\section[Approach to the infinite-$N$ limit]{\boldmath Approach to the infinite-$N$ limit}
\begin{figure}[ht]
\centerline{
\includegraphics[width=90mm]{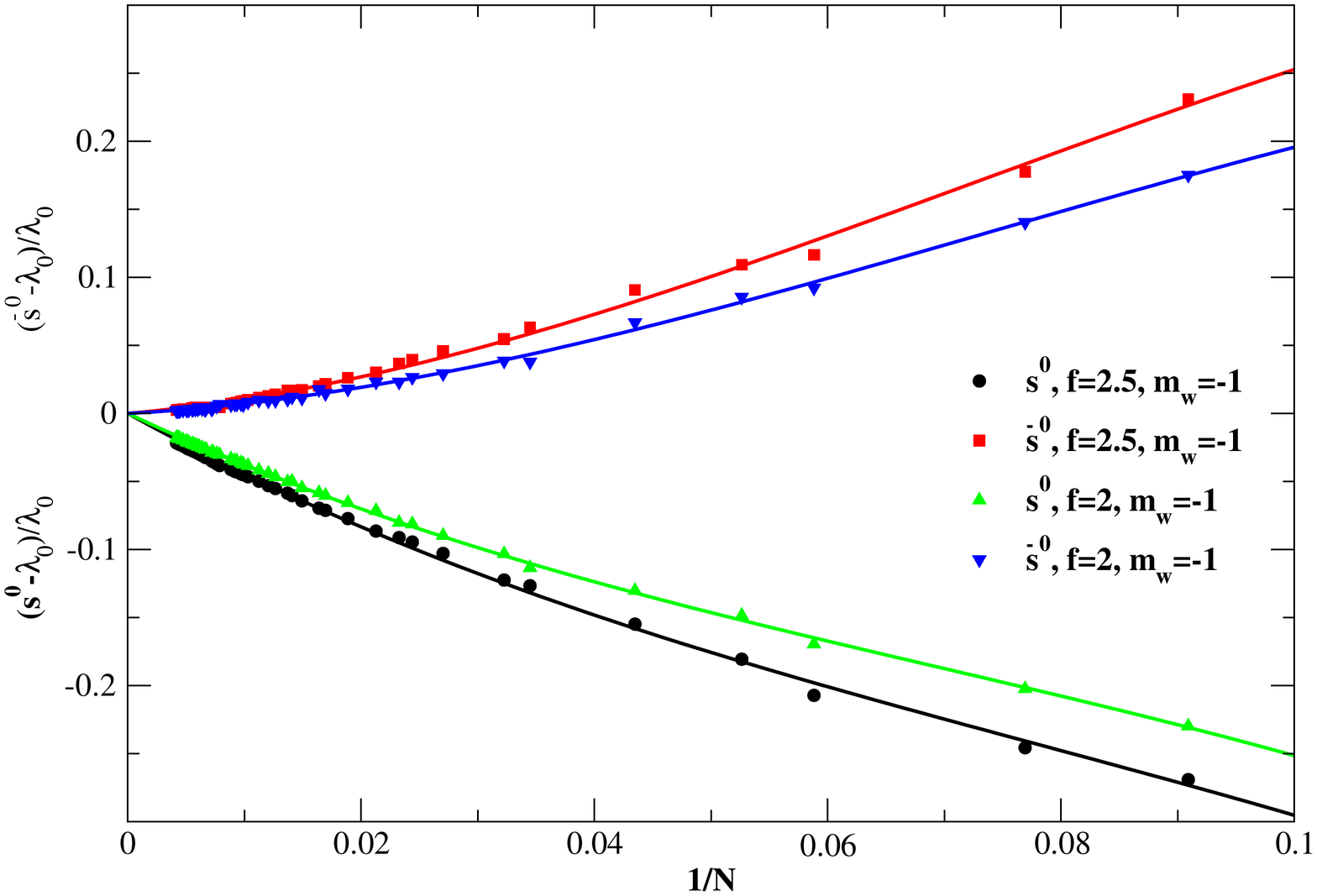}
\includegraphics[width=90mm]{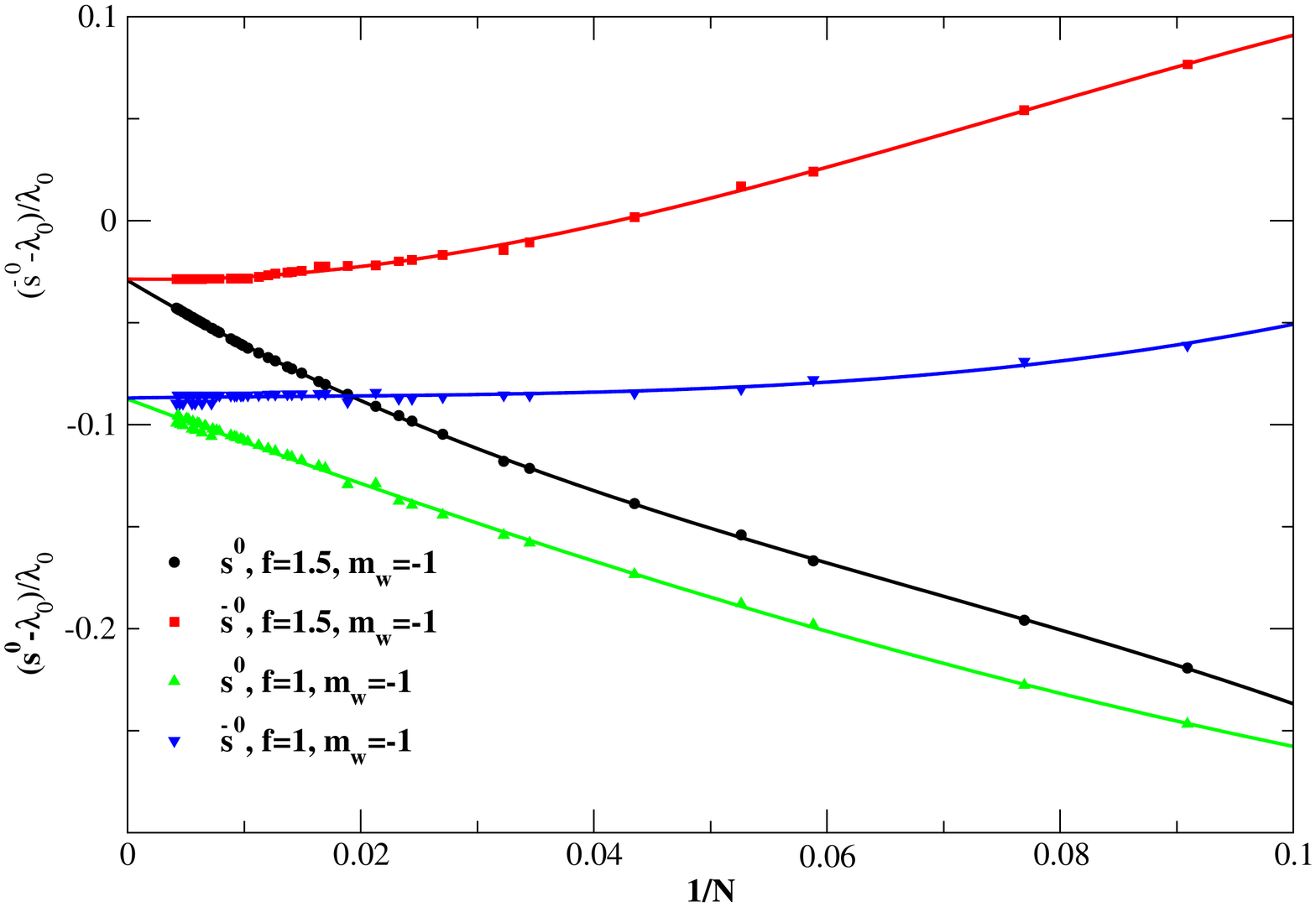}
}
\centerline{
\includegraphics[width=90mm]{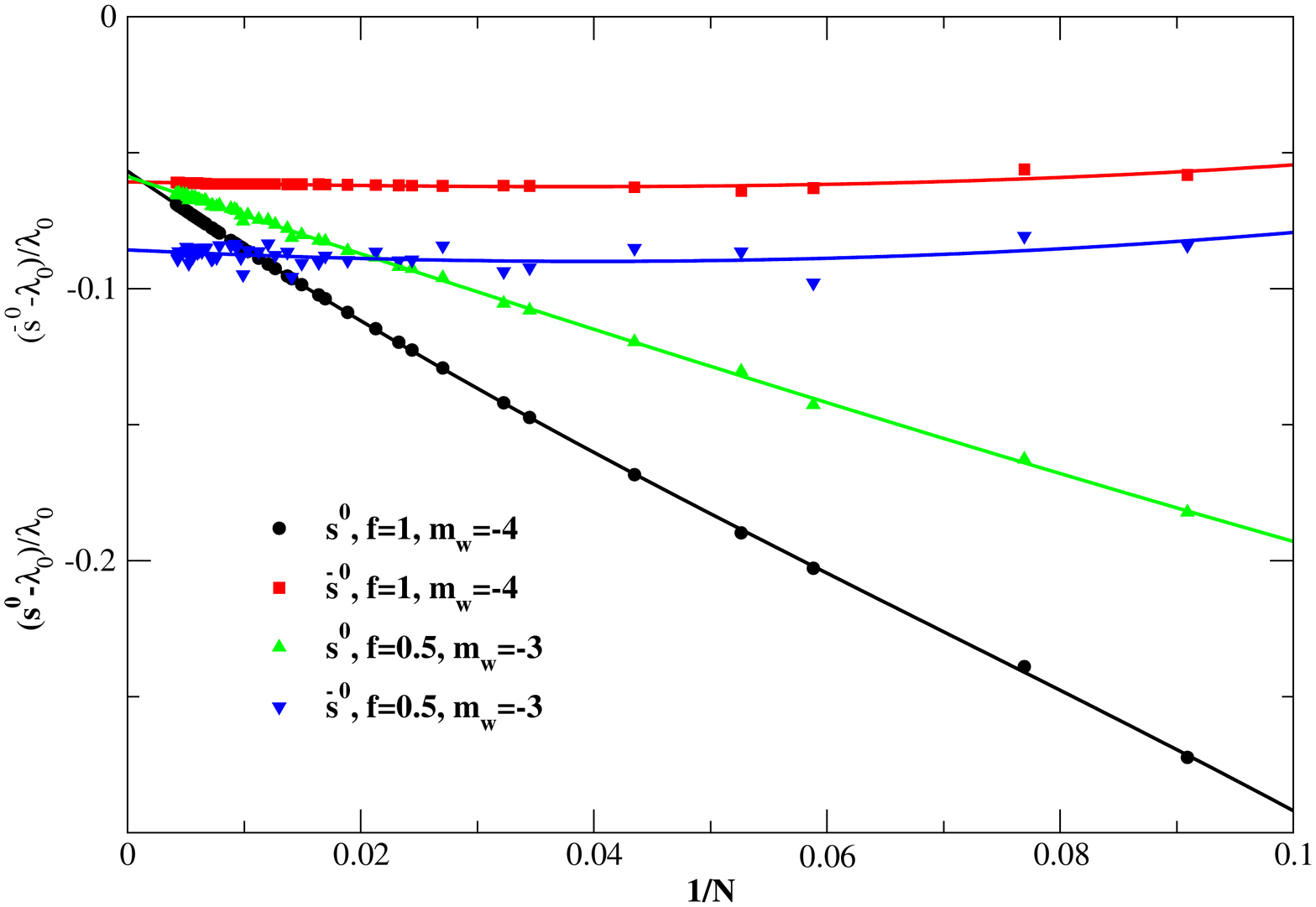}
}
\caption{
Approach of the lattice action density to the large-$N$ limit. The
left panel shows that we approach a uniform distribution in the
allowed region. 
The right panel shows that we do not approach a
uniform distribution in the region that is not allowed.
The bottom panel shows two other cases where we do not approach a
uniform distribution in the region that is not allowed (note that $\lambda_0<0$, i.e., $\lim_{N\to\infty}s^0>\lambda^0$).
} \label{action-plots}
\end{figure}

Given the leading-order partition function in (\ref{zquad}), we view the
action as a function of the $4N$ angles subject to the condition that
they belong to $\SU(N)$ and perform a maximization of the action using
the Hybrid Monte Carlo algorithm as described
in~\cite{Hietanen:2009ex}.
Instead of looking at the distribution of the angles which can look
uniform
due to the action being invariant under $\theta_\mu^i \to \theta_\mu^i
+ \frac{2\pi k_\mu}{N}$ for arbitrary integers $k_\mu$, we look at the
action $S^0$ and compare to what one
would get if we replace $\rho(\theta)$ by the constant distribution
$\frac{1}{(2\pi)^4}$ in (\ref{dentree}) which we will be $N^2\lambda_0$.
If the distributions at finite $N$ given by (\ref{distdef}) for the
maximum action configurations approach the uniform distribution as $N\to\infty$, we
expect
the action density, $s^0=\frac{S^0}{N^2}$, to approach
$\lambda_0$.
The plots in the top left panel of Fig.~\ref{action-plots} are for two
points in the allowed region and there is evidence for $s^0$
approaching $\lambda_0$ as $N\to\infty$. Contrary to this result, we
see that $s^0$ does not approach $\lambda_0$ for two points outside
the allowed region shown in the top right panel where we have only
changed
$f$ and kept $m_w=-1$ compared to the points shown in the top left
panel.
The bottom panel shows two other cases outside the allowed region
where the parameters coincide with previous numerical work~\cite{Hietanen:2012ma}.

\begin{figure}[ht]
\centerline{
\includegraphics[width=90mm]{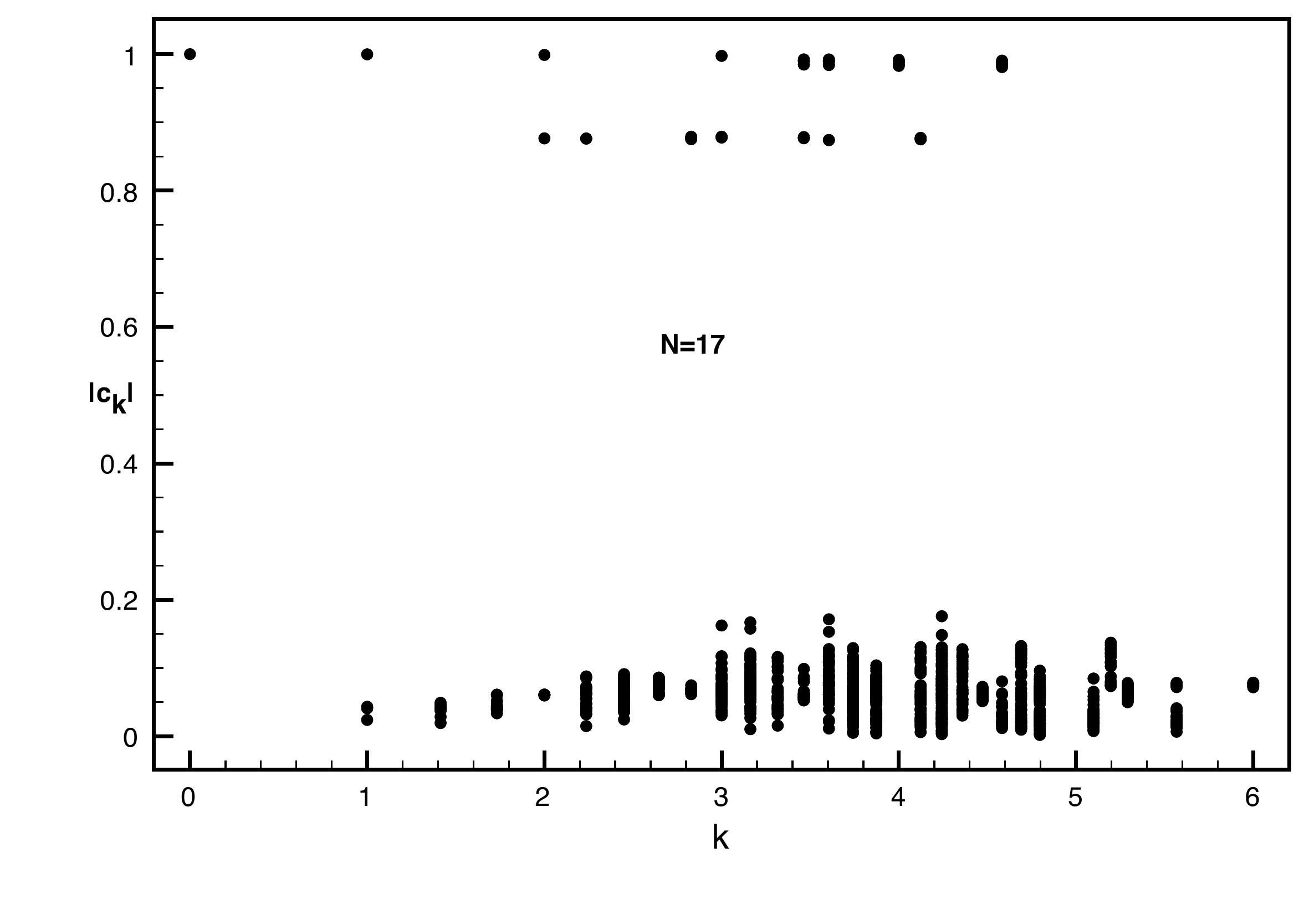}
\includegraphics[width=90mm]{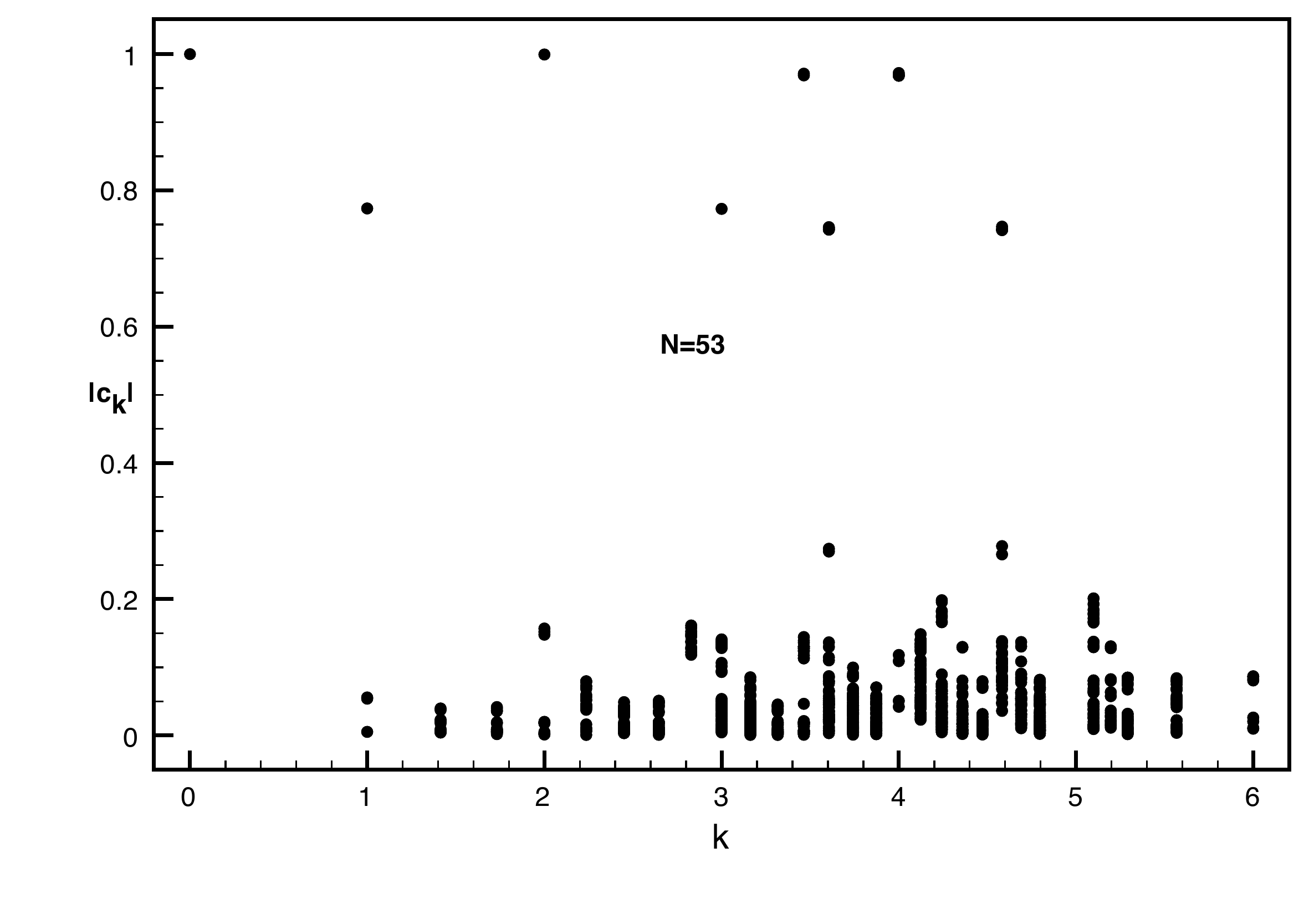}}
\centerline{
\includegraphics[width=90mm]{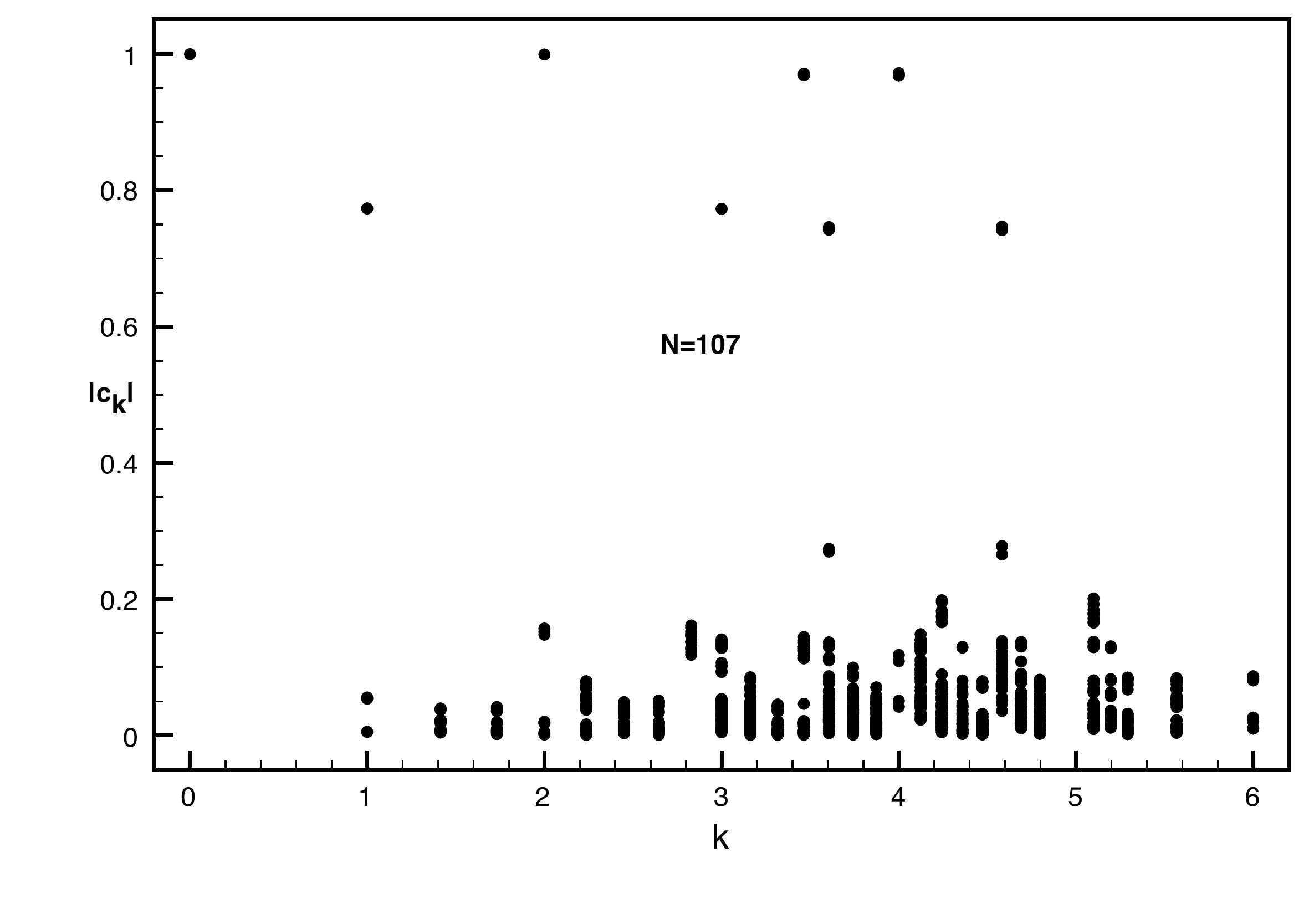}
\includegraphics[width=90mm]{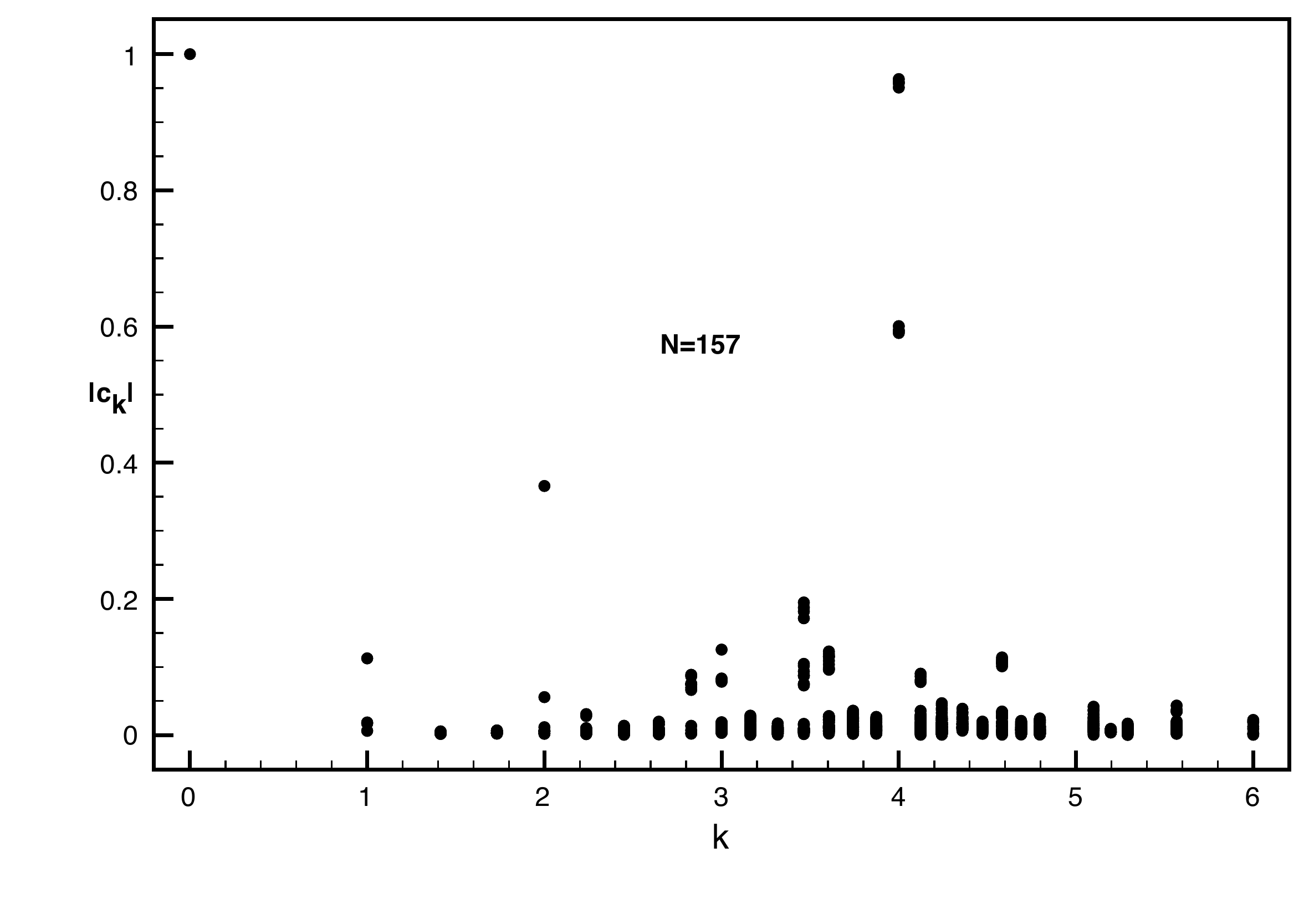}
}
\includegraphics[width=90mm]{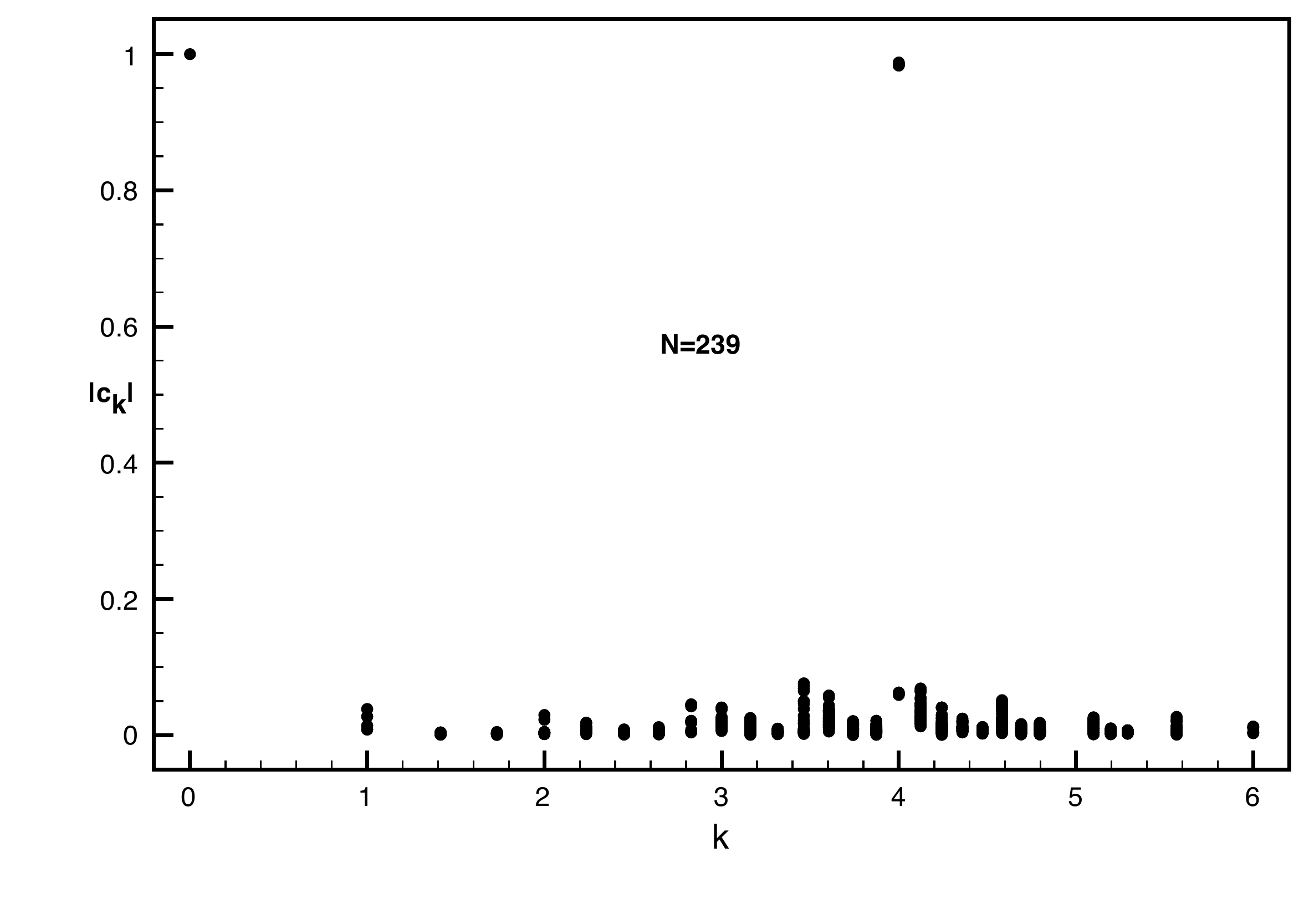}
\caption{
The Fourier coefficients, $|c_k|$, of $\rho(\theta)$ as a function of
$k=\sqrt{k^2}$ for several values of $N$ with $f=2$ flavors of massless overlap
fermions and the Wilson mass parameter set to $m_w=-1$.
} \label{fig8}
\end{figure}

\begin{figure}[ht]
\centerline{
\includegraphics[width=90mm]{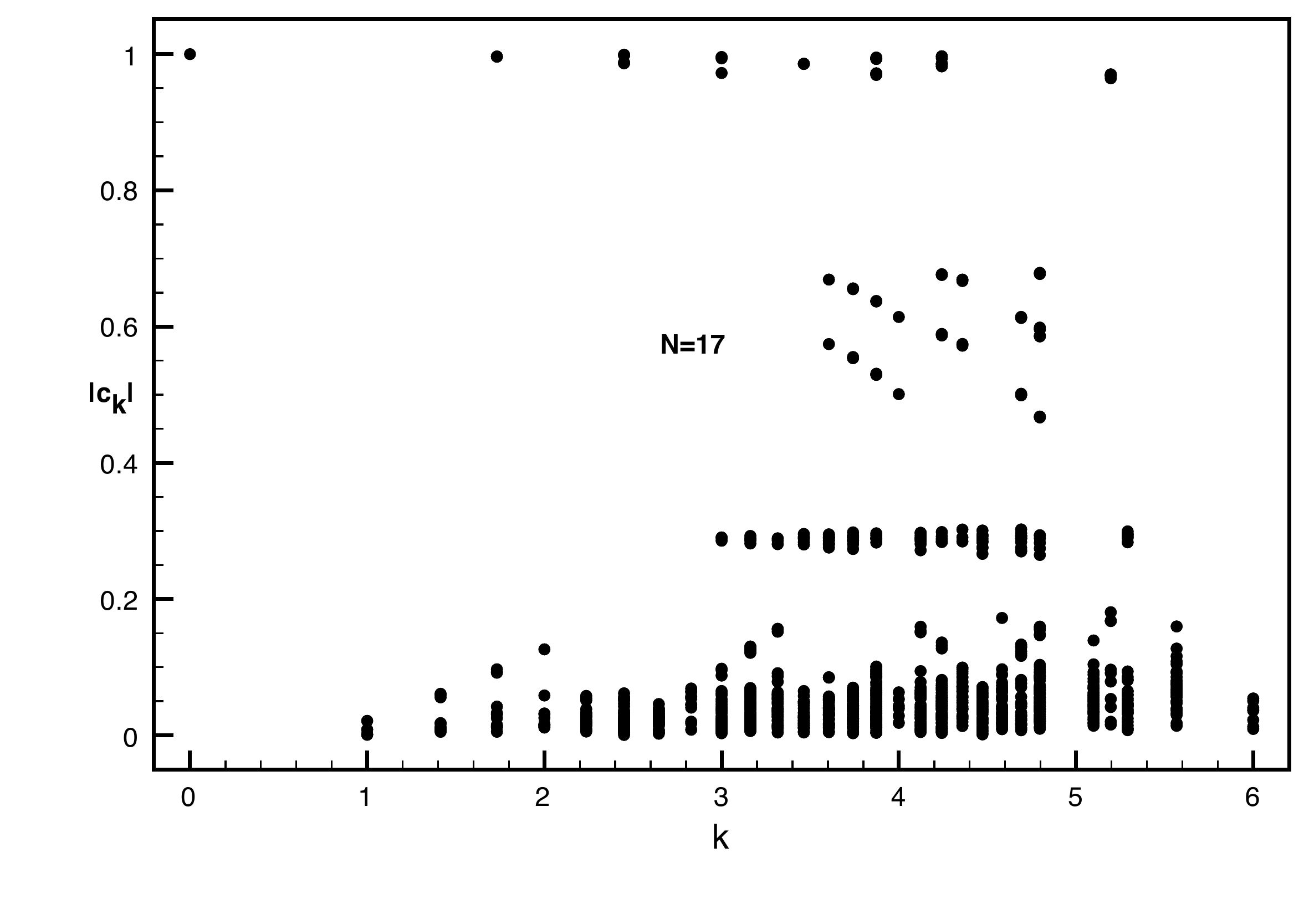}
\includegraphics[width=90mm]{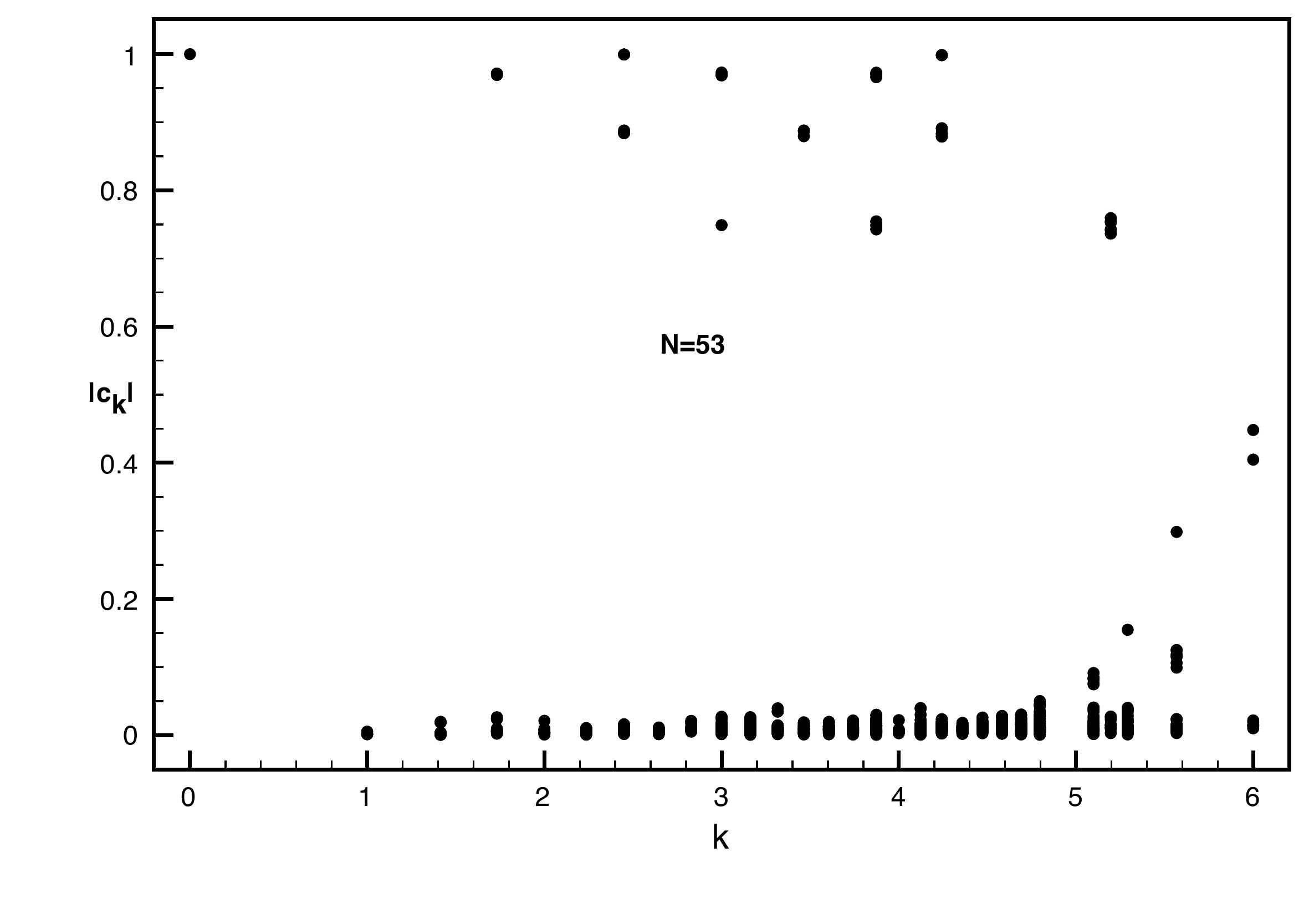}}
\centerline{
\includegraphics[width=90mm]{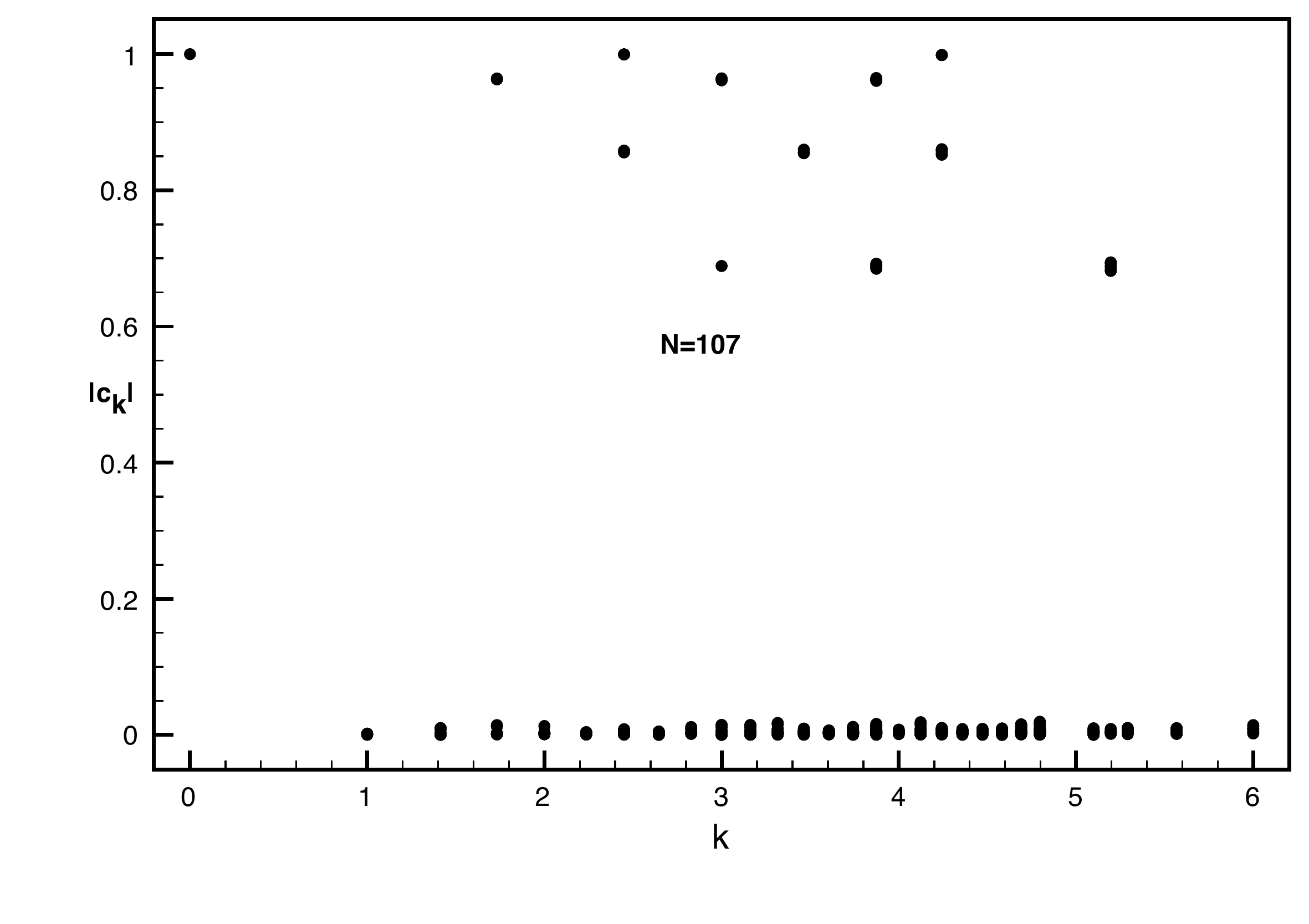}
\includegraphics[width=90mm]{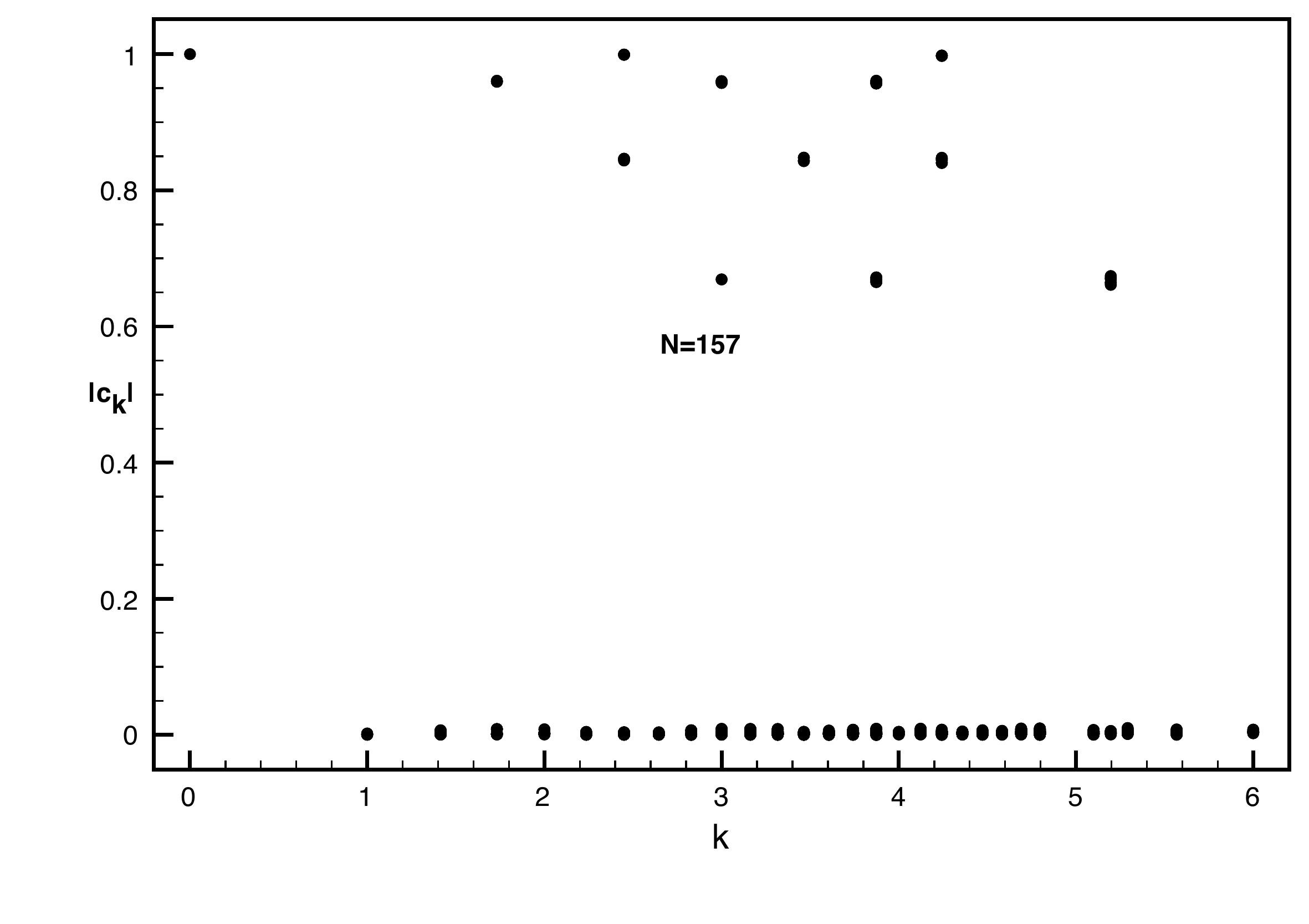}
}
\centerline{
\includegraphics[width=90mm]{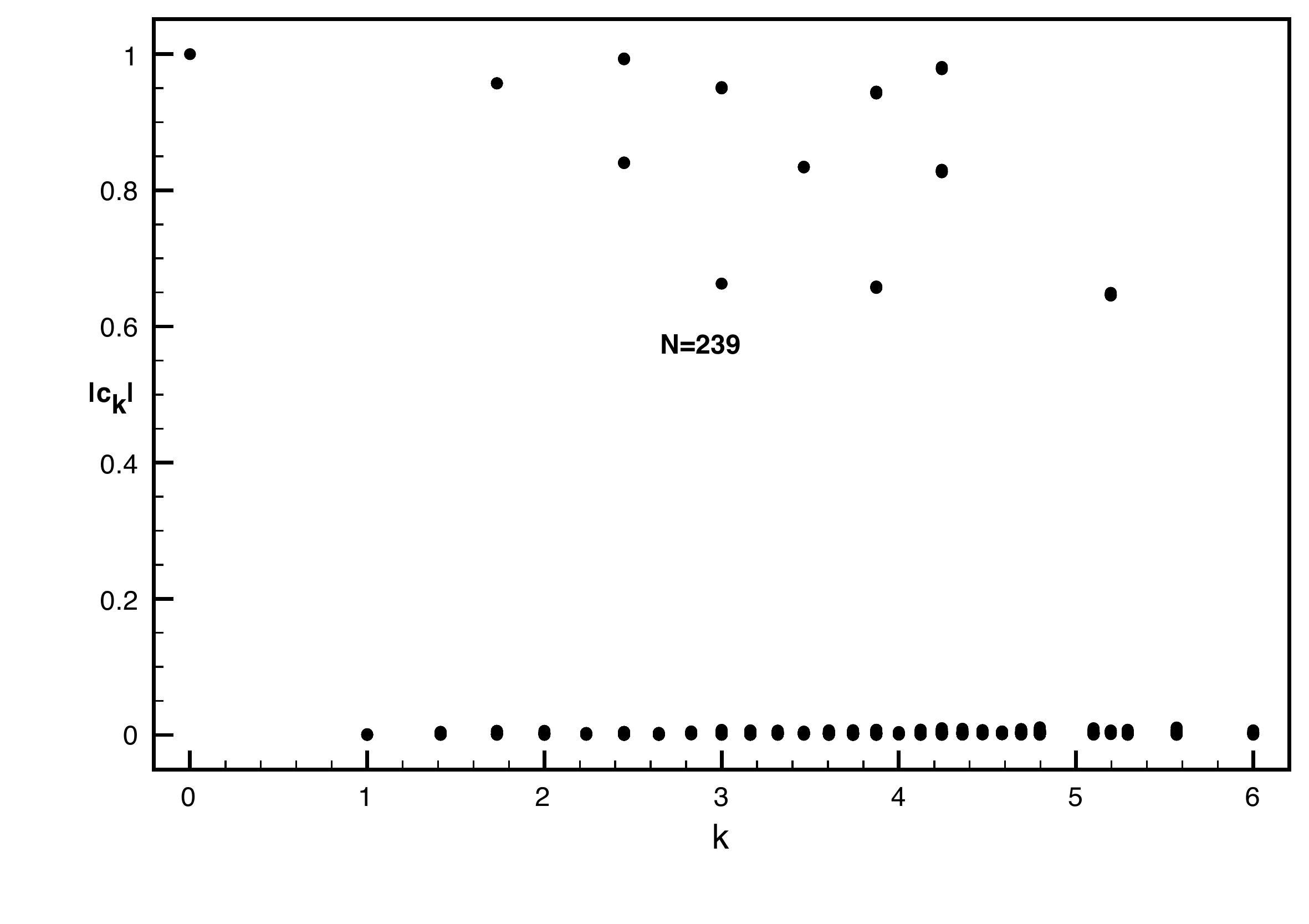}
\includegraphics[width=90mm]{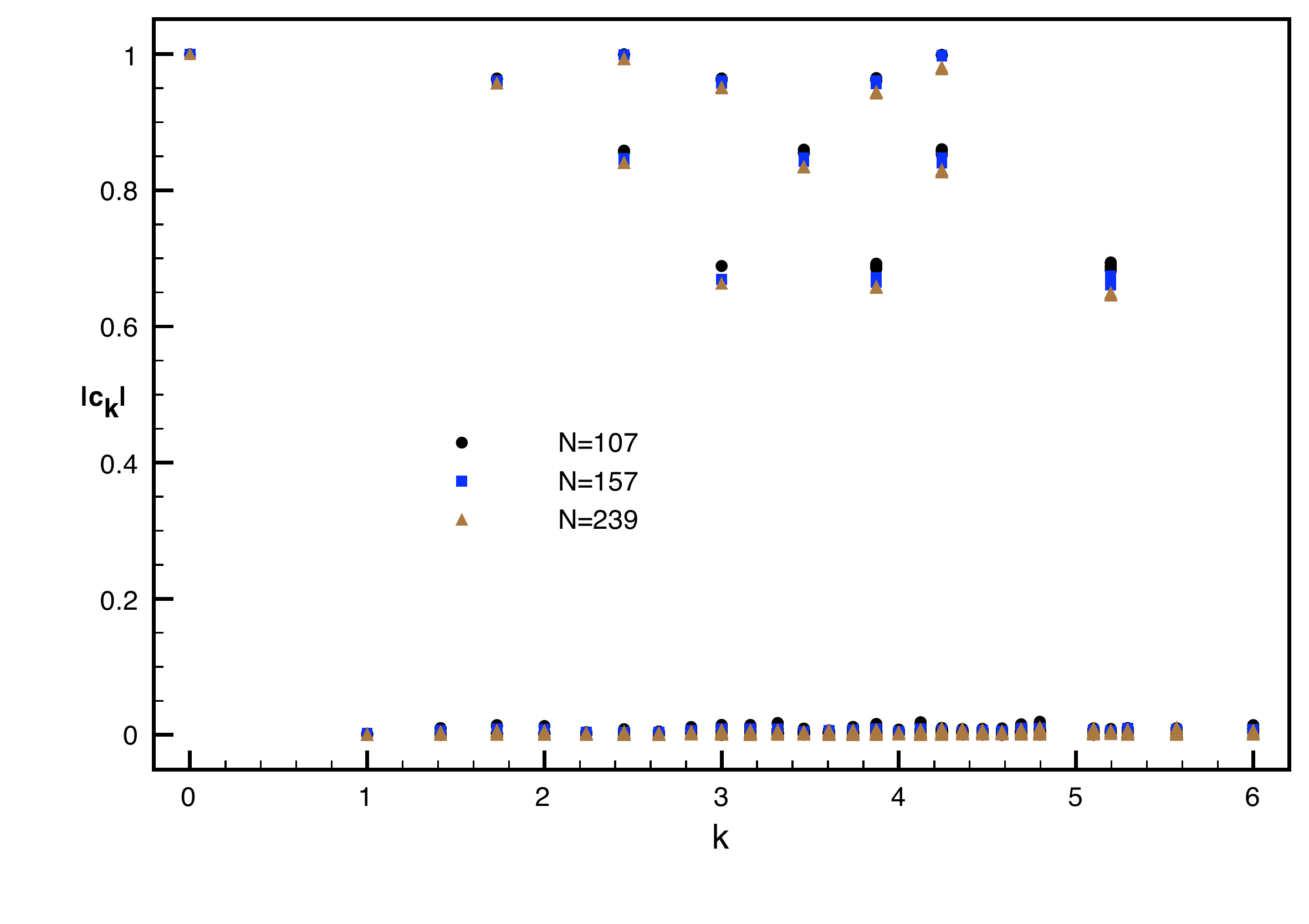}
}
\caption{
The Fourier coefficients, $|c_k|$, of $\rho(\theta)$ as a function of
$k=\sqrt{k^2}$ for several values of $N$ with $f=1$ flavors of massless overlap
fermions and the Wilson mass parameter set to $m_w=-4$.
} \label{fig9}
\end{figure}

In order to provide further support for the argument in the previous
paragraph,
we will start with the distribution at finite $N$ as given in
(\ref{distdef}) for the maximum action configuration $\{\theta_\mu^i\}$ and compute all
Fourier coefficients
\begin{align}
c_k = \int \prod_\mu d\theta_\mu \rho(\theta) e^{-i \sum_\nu
  k_\nu\theta_\nu} = \frac1N \sum_{j=1}^N e^{-i\sum_\nu k_\nu
  \theta_\nu^j} = \frac 1N \Tr\left(\prod_\nu
  D_\nu^{k_\nu}\right)^\dagger
\label{Fourierck}
\end{align}
for all $k$ with $-3 \le k_\mu \le 3$. The results for several values
of $N$ for one point in the allowed region ($f=2$, $m_o=0$, $m_w=-1$)
that coincides with a point in the top left panel of Fig.~\ref{action-plots} 
are shown in Fig.~\ref{fig8}. Results for one point outside the allowed
region ($f=1$, $m_o=0$, $m_w=-4$)
that coincides with a point in the bottom panel of Fig.~\ref{action-plots}
are shown in Fig.~\ref{fig9}. We expect all the Fourier
coefficients shown in Fig.~\ref{fig8} to approach zero and there is
some evidence for this. 
We can imagine constructing a sequence of distributions for $N=n^4$ (with $n=2,3,\ldots$) that approaches a uniform distribution in the large-$N$ limit by locating $\delta$-functions on all sites of a four-dimensional periodic hypercubic lattice with lattice spacing $2\pi/n$. The corresponding Fourier coefficients $c_k$ would by $1$ if all $k_\mu$ are multiplies of $n$ and zero otherwise. It is therefore not surprising that we obtain non-zero Fourier coefficients $c_k$ with $k$ being of order $N^{\frac 14}$ in Fig.~\ref{fig8} even though we expect all coefficients to vanish in the large-$N$ limit.

On the other hand, we expect some of the
Fourier coefficients shown in Fig.~\ref{fig9} to
approach a non-zero limit and there is some evidence for this
particularly when we look at 
the combined plot for $N=107, 157, 239$
shown in the bottom right panel.

If some of the Fourier coefficients shown in Fig.~\ref{fig9} indeed
approach a non-zero limit,
 the {\sl
  partial continuum action density} defined as
\begin{align}
\bar s^0 = \sum_{|k_\mu|\leq 3} c_k c_k^* \lambda_k\,,
\end{align}
should not approach $\lambda_0$. There is clear evidence for it when
we look at the plots in the top right and bottom panels of
Fig.~\ref{action-plots}.
Note that the approach to $N\to\infty$ is quite flat consistent with
the convergence seen in the combined plot for $N=107, 157, 239$
shown in the bottom right panel. Furthermore, the limit of $s^0$ and
$\bar s^0$ do not seem to coincide in the bottom panel of
Fig.~\ref{action-plots}
for the case of $f=0.5$ and $m_w=-3$ suggesting that there are modes
with
$k$ not in $-3 \le k_\mu \le 3$  that approach a non-zero limit at
infinite $N$.

Only in the limit of infinite $N$, we are allowed to ignore the
restriction of the principal value and  sum the
infinite series in (\ref{eq:S-ck}) to obtain a finite action density provided the
distribution has a smooth limit. The partial sum $\bar s^0$, on the
other hand, is finite at any $N$ but will only agree with $s^0$ at
infinite $N$ if all the coefficients not included in the sum approach
zero excluding some accidental cancellation due to eigenvalues with
different signs. If the distribution is uniform in the infinite-$N$
limit
as is expected for points in the allowed region, we expect $s^0$ and
$\bar s^0$ to coincide with $\lambda_0$. There is evidence for this in
the
top left panel of Fig.~\ref{action-plots}.

The computation of $\lambda_k$ in (\ref{eigen2}) should exclude a
small region of order $\epsilon$ around $\phi=0$ in order to properly
account for the principal value required at finite $N$ due to the form
of the distribution in (\ref{distdef}). One has to tune $\epsilon$ as
a function of $N$ and include a sum over all modes in (\ref{eq:S-ck})
(which will be finite) to match with $s^0$ at finite $N$. The
difference between $s^0$ and $\bar s^0$ at finite $N$ seen in
Fig.~\ref{action-plots}
is a combination of two effects: not excluding a small region of order
$\epsilon$
and not including a sum over all modes.

\section {Discussion of previous numerical work}

Previous numerical work described in~\cite{Hietanen:2009ex} only
looked at
one Fourier mode, namely, $k_\mu=(1,0,0,0)$ and its permutations.
We know from the analysis of the Fourier modes in this paper, that
some
coefficients could be accidentally small and it is necessary to look
at
several Fourier modes.
The wilson mass parameter was 
set to $m_w=-5$ in the numerical analysis performed
in~\cite{Hietanen:2009ex} at finite lattice coupling
and we know from the analysis performed here that this is 
not in the allowed region for any value of $f$ in the weak-coupling limit.

Numerical work in~\cite{Hietanen:2012ma} falls under a slightly different
category. The running of the coupling studied in that paper at the
range
of lattice couplings did not agree with two loop perturbation theory.
Since the theory studied used massless overlap fermions with $f=1$ and
$m_w=-4$,
we know from the analysis performed here that we cannot obtain an
infinite-volume continuum limit by going to $b\to\infty$ (weak-coupling limit).
The speculative part in~\cite{Hietanen:2012ma} suggests the possibility of
a continuum limit away from $b=\infty$. If this is the case, then the
analysis performed in this paper does not shed light into such a
scenario.

\begin{figure}[ht]
\centerline{
\includegraphics[width=140mm]{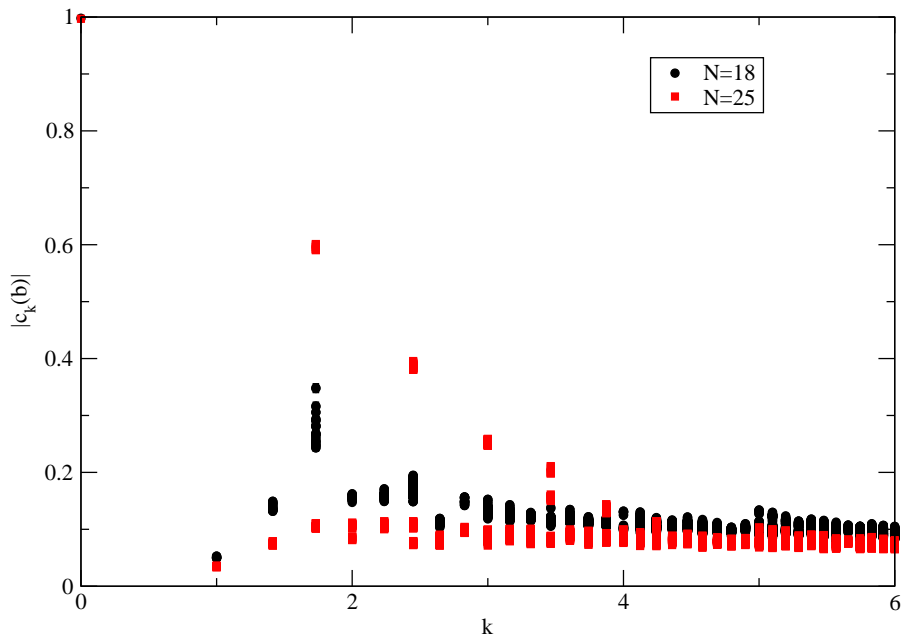}
}
\caption{
The Fourier coefficents, $c_k(b),$ as a
function of $k=\sqrt{k^2}$ at finite lattice coupling, 
$b=0.65,$ with one flavor of massless overlap Dirac fermions and $m_w=-4$.
} \label{ckb}
\end{figure}

Numerical studies with two flavors of Wilson fermions (both massless
and massive) were carried out in~\cite{Bringoltz:2011by}. 
From Fig.~\ref{fig:wilson-f-vs-mw} in 
this paper, we know that we cannot obtain the infinite-volume 
continuum limit with 
one or two flavors of Wilson fermions. 
Evidence for
being in the correct continuum phase was obtained by a study of
operators
of the form $|\Tr U_\mu|$, $|\Tr U_\mu U_\nu|$ and $|\Tr U_\mu
U_\nu^\dagger|$.
These can be considered as special cases of 
\begin{align}
c_k(b) = \Tr U_1^{k_1} U_2^{k_2} U_3^{k_3} U_4^{k_4}
\end{align}
which will tend to $c_k$ in (\ref{Fourierck}) in the weak-coupling
limit, $b\to\infty$. The ordering of the operators will matter at
finite lattice coupling if more than two of the $k_\mu$ are non-zero. 
We do not have the gauge field data for Wilson fermions
used
in~\cite{Bringoltz:2011by} but we do have them for massless overlap fermions at
$b=0.65$, $f=1$ and $m_w=-4$ in~\cite{Hietanen:2012ma}. 
The results for $c_k(b)$ are plotted in
Fig.~\ref{ckb}
and should be compared with Fig.~\ref{fig9}. We see that $b=0.65$ is
far away from the weak-coupling limit consistent with the results in~\cite{Hietanen:2012ma}.
Furthermore, the coefficients with small $k$ could be accidentally
small. 
This suggests that the conclusions
in~\cite{Bringoltz:2011by}
possibly result from being far away from the weak-coupling limit and
not looking at a sufficient number of Fourier modes.

In addition, we can also conclude that we cannot use a single-site
model with heavy Wilson fermions and obtain the continuum limit of a
pure gauge theory in contrast to the claims made in~\cite{Hanada:2013ota}.
The eigenvalues of Wilson fermions are
doubly
degenerate~\cite{Hietanen:2009ex} but they do not come in pairs of
opposite
chirality. Therefore, it is not apriori clear how to deal with half a
flavor of Wilson fermions making it essentially impossible to study
the
continuum limit of any infinite-volume theory using adjoint Wilson
fermions on a single-site lattice.

Since we cannot keep the bare mass finite and non-zero as we take the
weak-coupling limit, the
proposal in~\cite{Azeyanagi:2010ne} to use single-site models with
massive
adjoint fermions in order to extract physics of pure gauge theories is
ruled out.

\section{Future Work}

The allowed regions plotted in Fig~\ref{fig:allowed} provide for
an interesting scenario when it comes to the usefulness of single-site
theories
to describe correct infinite-volume continuum physics. We cannot study
theories with $f\le \frac{3}{2}$ unless we entertain the possibility
that
the continuum limit occurs away from $b=\infty$ but this would be a
radical
deviation from conventional wisdom. We can study $f\ge 2$ theories
without a fermion mass. It is possible we can study $f\ge 2$ theories
with a fermion mass provided we can stay in the correct phase by
taking the bare mass to zero as we go to the weak-coupling limit such
that
the physical mass is kept constant. Since this theory is expected to be
conformal
for massless fermions, it is not clear how a single-site theory will
exhibit conformal behavior. This is certainly a case worth further
investigation.

\begin{acknowledgments} 
The authors acknowledge partial support by the NSF under grant numbers
PHY-0854744 and PHY-1205396.
\end{acknowledgments}

\end{document}